\documentclass[12pt,reqno]{amsart}

\usepackage{amsmath,amsfonts,amsthm,cite,amssymb}

\usepackage{times}
\usepackage{a4wide}
\usepackage{fullpage}
\usepackage[footnotesize,hang,sf]{caption}

\usepackage{titlesec}
\titleformat{\section}
{\bfseries\scshape\centering}
{\thesection.}{.5em}{}
\titleformat{\subsection}
{\rmfamily\bfseries}
{\thesubsection}{.5em}{}
                                                                                
\usepackage{psfrag}
\usepackage[dvips]{graphicx}
\usepackage{pstricks}
\usepackage{pst-3dplot}
\numberwithin{equation}{section}

\newtheorem{theorem}{Theorem}
\newtheorem{prop}[theorem]{Proposition}
\newtheorem{coro}[theorem]{Corollary}

\DeclareMathOperator{\sign}{sign}
\DeclareMathOperator{\Tr}{Tr}
\DeclareMathOperator{\sh}{sh}
\DeclareMathOperator{\ch}{ch}

\DeclareMathOperator{\CS}{CS}

\DeclareMathOperator{\erfc}{erfc}

\begin{document}

\renewcommand{\thefootnote}{\fnsymbol{footnote}}
\baselineskip 17pt
\parskip 7pt
\sloppy




\title{On the Quantum  Invariant for the Brieskorn Homology Spheres}


    \author{Kazuhiro \textsc{Hikami}}


  \address{Department of Physics, Graduate School of Science,
    University of Tokyo,
    Hongo 7--3--1, Bunkyo, Tokyo 113--0033,   Japan.
    }
    
    \urladdr{http://gogh.phys.s.u-tokyo.ac.jp/{\textasciitilde}hikami/}

    \email{\texttt{hikami@phys.s.u-tokyo.ac.jp}}


\date{May 2, 2004. Accepted on  December 21, 2004.}

\begin{abstract}
We study an exact asymptotic behavior of the
Witten--Reshetikhin--Turaev SU(2) invariant for the
Brieskorn homology spheres $\Sigma(p_1,p_2,p_3)$
by use of  properties of the modular form following
a method proposed by
R.~Lawrence and D.~Zagier.
Key observation is  that the invariant coincides with a limiting value of the
Eichler integral of the modular form with weight $3/2$.
We show that the Casson invariant is related to the number
of the  Eichler integrals which do not vanish  in a limit
$\tau\to N \in \mathbb{Z}$.
Correspondingly  there is a one-to-one correspondence
between the non-vanishing Eichler integrals and
the irreducible representation of the fundamental group,
and the Chern--Simons invariant is  given from the
Eichler integral in  this limit.
It is also shown that the Ohtsuki invariant follows from a nearly modular property of the Eichler integral, and we give an explicit form in terms of the $L$-function.

\end{abstract}





\maketitle

\section{Introduction}

The quantum invariant for the 3-manifold $\mathcal{M}$
was  introduced  as a path integral on $\mathcal{M}$ by
Witten~\cite{EWitt89a};
as the $SU(2)$ invariant we have
\begin{equation}
  \label{Witten}
  Z_{k} (\mathcal{M})
  =
  \int \exp\Bigl(
    2 \, \pi \, \mathrm{i} \, k \,
    \CS(A)
  \Bigr) \,
  \mathcal{D} A
\end{equation}
where $k \in \mathbb{Z}$, and
$\CS(A)$ is the Chern--Simons functional defined by
\begin{equation}
  \CS(A)=
  \frac{1}{8 \, \pi^2} \,
  \int\limits_{\mathcal{M}}
  \Tr
  \left(
    A \wedge \mathrm{d} A +
    \frac{2}{3} \, A \wedge A \wedge A
  \right)
\end{equation}
Since this work,
studies of the quantum invariants of the 3-manifolds have
been extensively
developed,
and  a construction of the 3-manifold invariant
was reformulated combinatorially and rigorously in
Refs.~\citen{ResheTurae91a,KirbMelv91a}
using a surgery description of $\mathcal{M}$ and the colored Jones polynomial defined in Ref.~\citen{Jones85}.

By applying a stationary phase approximation,
an asymptotic behavior of the Witten invariant
in large $k\to\infty$
is expected to be~\cite{EWitt89a,FreeGomp91a}
(see also Ref.~\citen{Atiya90Book})
\begin{equation}
  \label{asymptotic_Witten}
  Z_k(\mathcal{M})
  \sim
  \frac{1}{2} \, \mathrm{e}^{-\frac{3}{4} \pi \mathrm{i}} \,
  \sum_\alpha
  \sqrt{T_\alpha(\mathcal{M})} \,
  \mathrm{e}^{- 2 \pi \mathrm{i} I_\alpha/4} \,
  \mathrm{e}^{2 \pi \mathrm{i} (k+2) \CS(A)}
\end{equation}
Here the sum runs over a flat connection $\alpha$, and
$T_\alpha$ and $I_\alpha$
are  the Reidemeister torsion
and the spectral flow defined modulo $8$ respectively.

\begin{figure}[htbp]
  \centering
  
    \begin{psfrags}
    \psfrag{x}{$0$}
    \psfrag{a}{$p_1/q_1$}
    \psfrag{b}{$p_2/q_2$}
    \psfrag{c}{$p_3/q_3$}
    \includegraphics[scale=1.0, bb=-60 -60 60 40]{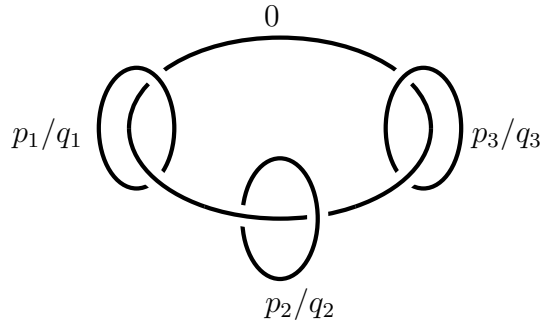}
  \end{psfrags}
  \caption{
    Rational surgery description of the Brieskorn homology sphere
    $\Sigma(p_1, p_2, p_3)$}
  \label{fig:Brieskorn}
\end{figure}

In this article
we consider the Brieskorn homology spheres $\Sigma(p_1, p_2, p_3)$
where $p_i$ are pairwise coprime  positive integers.
This is the intersection of the singular complex surface
\begin{equation*}
  z_1^{~p_1} +   z_2^{~p_2} +   z_3^{~p_3} =0
\end{equation*}
in complex three-space with the unit five-sphere
$|z_1|^2 + |z_2|^2  + |z_3|^2  =1$.
The manifold $\Sigma(2,3,5)$ is the Poincar{\'e} homology sphere.
These manifolds have a rational surgery description as in
Fig.~\ref{fig:Brieskorn}, and the fundamental group has the
presentation
\begin{equation}
  \label{presentation}
  \pi_1 \bigl( \Sigma(p_1,p_2,p_3) \bigr)
  =
  \Bigl\langle
  x_1, x_2, x_3, h
  ~\big|~
  \text{$h$  center},
  \text{$x_k^{~p_k} = h^{-q_k}$ for $k=1,2,3$},
  x_1 \, x_2 \, x_3 =1
  \Bigr\rangle
\end{equation}
where $q_k \in \mathbb{Z}$ such that
\begin{gather}
  \label{P_p_q}
  P \,  \sum_{k=1}^3 \frac{q_k}{p_k} = 1
\end{gather}
Here and hereafter we use
\begin{gather}
  P = P(p_1,p_2, p_3)
  = p_1 \, p_2 \, p_3
\end{gather}

Asymptotic behavior of the quantum invariant for the homology spheres was
studied in
Refs.~\citen{Rozan95a,Rozan96c,Rozan96d,Rozan96e,LRozan96f,LRozan97a,RLawre95a,RLawre96a,LawreRozan99a}.
Our purpose here  is to reformulate these results
number theoretically by use of properties of the
modular form following a method  of Lawrence and
Zagier~\cite{LawrZagi99a}.
A key observation is a fact that the Witten--Reshetikhin--Turaev (WRT)
invariant for the Brieskorn homology spheres is regarded as a limit
value of the Eichler integral of the modular form with weight $3/2$ as
was suggested in Ref.~\citen{LawrZagi99a}.
Using a nearly modular property of the Eichler integral, we can derive
an exact asymptotic behavior of the WRT invariant.
Correspondingly,
we can find an interpretation for topological  invariants such as the
Chern--Simons invariant,
the Casson invariant~\cite{KWalk92a},
the Ohtsuki invariant~\cite{HMuraka93a,HMuraka95a,Ohtsu95b,TOhtsu96a},
and the Ray--Singer--Reidemeister torsion
from the viewpoint of the modular form.

This paper is organized as follows.
In section~\ref{sec:invariant} we construct the WRT
invariant for the Brieskorn homology spheres
$\Sigma(p_1,p_2,p_3)$
following Ref.~\citen{LawreRozan99a}.
We use a surgery description of the 3-manifold, and apply
a formula in Ref.~\citen{LCJeff92a}.
In section~\ref{sec:modular} we introduce the modular form with weight $3/2$.
This gives
a $(p_1-1) \, (p_2 -1 ) \, (p_3 -1)/4$-dimensional representation of the
modular group $PSL(2;\mathbb{Z})$.
We consider the Eichler integral thereof in section~\ref{sec:Eichler}.
We study a limiting value of the Eichler integral at $\tau\in \mathbb{Q}$,
and show that the number of the Eichler integrals
which have non-zero value in a limit
$\tau\to N\in\mathbb{Z}$
is related to the Casson invariant of the Brieskorn homology spheres.
It will be discussed 
that we have a one-to-one correspondence with the irreducible representation of the fundamental group,
and that the Chern--Simons invariant is given from this
limiting value of the Eichler integral.
We further give a nearly modular property of the Eichler integral.
In section~\ref{sec:WRT}
we reveal a key identity that the WRT invariant for the Brieskorn spheres coincides with
a limiting value of the Eichler integral at $\tau\to 1/N$.
This was suggested in Ref.~\citen{LawrZagi99a}
where  a correspondence was proved only
for a case of the Poincar{\'e} homology sphere.
Combining with  results in section~\ref{sec:Eichler} we obtain an exact
asymptotic expansion of the WRT invariant
which is constituted from two terms;
one is a sum of dominating exponential term which gives the Chern--Simons term,
and another  is a ``tail''  part which may be regarded as a contribution from a trivial connection.
In section~\ref{sec:torsion} we prove that the $\mathbf{S}$-matrix
of the modular transformation gives 
both the Reidemeister torsion and
the spectral flow.
We show in section~\ref{sec:Ohtsuki} that a tail part gives the Ohtsuki invariant.
Explicitly computed is the $n$-th Ohtsuki invariant in terms of the $L$-function.
The last section is devoted to  concluding remarks.

\section{The Witten--Reshetikhin--Turaev Invariant for the Brieskorn
  Homology Spheres}
\label{sec:invariant}

We introduce  the Reshetikhin--Turaev
invariant $\tau_N(\mathcal{M})$~\cite{ResheTurae91a} for $N\in\mathbb{Z}_+$.
This  is related to the Witten invariant
$Z_k(\mathcal{M})$ defined in eq.~\eqref{Witten} as
\begin{equation}
  \label{Witten_and_tau}
  Z_k(\mathcal{M})
  =
  \frac{
    \tau_{k+2}(\mathcal{M})
  }{
    \tau_{k+2}(S^2 \times S^1)
  }
\end{equation}
Here  the invariant is normalized to be
\begin{gather*}
  \tau_N (S^3) =1
\end{gather*}
and we have
\begin{gather*}
  \tau_N(S^2 \times S^1)
  =
  \sqrt{\frac{N}{2}} \, \frac{1}{\sin(\pi/N)}
\end{gather*}


When the 3-manifold $\mathcal{M}$ is constructed by the rational
surgeries
$p_j / q_j$
on the $j$-th component of $n$-component link $\mathcal{L}$, it was
shown~\cite{ResheTurae91a,LCJeff92a} that the invariant $\tau_N(\mathcal{M})$ is
given by
\begin{equation}
  \label{define_invariant}
  \tau_N(\mathcal{M})
  =
  \mathrm{e}^{
    \frac{\pi \mathrm{i}}{4} \frac{N-2}{N}
    \left(
      \sum_{j=1}^n \Phi(U^{(p_j,q_j)})
      -3 \sign(\mathbf{L})
    \right)
  }
  \sum_{k_1, \dots, k_n=1}^{N-1}
  J_{k_1,\dots,  k_n}(\mathcal{L}) \,
  \prod_{j=1}^n
  \rho(U^{(p_j,q_j)})_{k_j, 1}
\end{equation}
Here the surgery $p_j/q_j$  is described by an $SL(2; \mathbb{Z})$ matrix
\begin{equation*}
  U^{(p_j,q_j)}=
  \begin{pmatrix}
    p_j & r_j \\
    q_j & s_j
  \end{pmatrix}
\end{equation*}
and $\Phi(U)$ is the Rademacher $\Phi$-function defined by
(see, \emph{e.g.}, Ref.~\citen{HRadema73})
\begin{equation}
  \Phi
  \left[
    \begin{pmatrix}
      p & r \\
      q & s
    \end{pmatrix}
  \right]
  =
  \begin{cases}
    \displaystyle
    \frac{p+s}{q} - 12 \, s(p, q) &
    \text{for $q \neq 0$}
    \\[3mm]
    \displaystyle
    \frac{r}{s}
    &
    \text{for $q=0$}
  \end{cases}
\end{equation}
where   $s(b,a)$ denotes the Dedekind sum  
(see, \emph{e.g.},  Ref.~\citen{RademGross72}, and
also Ref.~\citen{KirbMelv94a})
\begin{align}
  \label{Dedekind_sum}
  s(b,a)
  & =
  \sign(a)
  \sum_{k=1}^{|a|-1}
  \Bigl(\Bigl( \frac{k}{a} \Bigr) \Bigr) \cdot
  \Bigl(\Bigl( \frac{k \, b}{a} \Bigr) \Bigr) 
\end{align}
with
\begin{equation*}
  ((x))
  =
  \begin{cases}
    x - \lfloor x \rfloor - \frac{1}{2}
    &
    \text{if $x \not\in \mathbb{Z}$}
    \\[3mm]
    0
    &
    \text{if $x \in \mathbb{Z}$}
  \end{cases}
\end{equation*}
and $\lfloor x \rfloor$ is the greatest integer not exceeding $x$.
It is known that the Dedekind sum is rewritten as
\begin{equation*}
  s(b,a)
  =
  \frac{1}{4 \, a}
  \sum_{k=1}^{a-1}
  \cot \left( \frac{ k}{a}  \, \pi \right) \,
  \cot \left( \frac{ k \, b }{a} \, \pi \right) 
\end{equation*}
An $n\times n$ matrix $\mathbf{L}$ is a linking matrix
$\mathbf{L}_{j,k} = \mathrm{lk}_{j,k} + {p_j}/{q_j} \cdot \delta_{j,k}$,
and $\sign(\mathbf{L})$ is a signature of $\mathbf{L}$, \emph{i.e.},
the
difference between the number of positive and negative eigenvalues of
$\mathbf{L}$.
The  polynomial $J_{k_1,\dots,k_n}(\mathcal{L})$ is the colored Jone
polynomial for link $\mathcal{L}$ with the color $k_j$ for the $j$-th
component link, and
$\rho(U^{(p,q)})$ 
is a representation $\rho$ of $PSL(2;\mathbb{Z})$;
\begin{multline}
  \rho(U^{(p,q)})_{a,b}
  \\
  =
  - \mathrm{i} \frac{\sign(q)}{\sqrt{2 \, N \, | q|}} \,
  \mathrm{e}^{
    -\frac{\pi \mathrm{i}}{4} \, \Phi(U^{(p,q)})
  } \,
  \mathrm{e}^{
    \frac{\pi \mathrm{i}}{2 N q} s b^2 
  } \,
  \sum_{
    \substack{
      \gamma \mod 2 N q
      \\
      \gamma = a \mod 2 N}
  }
  \mathrm{e}^{
    \frac{\pi \mathrm{i}}{2 N q}
    p \gamma^2
  } \,
  \left(
    \mathrm{e}^{
      \frac{\pi \mathrm{i}}{ N q}
       \gamma b
    }
    -
    \mathrm{e}^{
      - \frac{\pi \mathrm{i}}{ N q}
      \gamma b
    }
  \right)
\end{multline}
for $1 \leq a, b \leq N-1$~\cite{LCJeff92a},
and we have
\begin{equation}
  \label{affine_su}
  \begin{gathered}
    \rho(S)_{a,b}
    =
    \sqrt{\frac{2}{N}} \,
    \sin
    \left( \frac{ a \, b \, \pi}{N} \right)
    \\[2mm]
    \rho(T)_{a,b}
    =
    \mathrm{e}^{\frac{\pi \mathrm{i}}{2 N} a^2 - \frac{\pi \mathrm{i}}{4}}
    \,
    \delta_{a,b}
  \end{gathered}
\end{equation}
with
\begin{align}
  \label{PSL_ST}
  S
  &=
  \begin{pmatrix}
    0 & -1 \\
    1 & 0
  \end{pmatrix}
  &
  T 
  & =
  \begin{pmatrix}
    1 & 1 \\
    0 & 1
  \end{pmatrix}
\end{align}
satisfying
\begin{equation*}
  S^2 = (S \, T)^3 = 1
\end{equation*}

We should note~\cite{RademGross72} that the Dedekind sum satisfies
\begin{gather}
  s(-b, a )= - s (b,a)
  \\[2mm]
  s(b,a) = s(b^\prime, a) \quad
  \text{for $b \, b^\prime \equiv 1 \pmod{a}$
  }
\end{gather}
and that the Rademacher $\Phi$-function fulfills
\begin{equation}
  \Phi
  (S \,  U^{(p, q)})
  =
  \Phi
  (  U^{(p, q)})
  - 3 \, \sign(p \, q)
\end{equation}

\begin{prop}[\cite{LawreRozan99a}]
  The WRT invariant for the Brieskorn homology spheres is given by
  \begin{multline}
    \label{Rozansky}
    \mathrm{e}^{
      \frac{2 \pi \mathrm{i}}{N}
      ( \frac{\phi}{4} - \frac{1}{2} )
    } \,
    \left(
      \mathrm{e}^{\frac{2 \pi \mathrm{i}}{N}} - 1
    \right)
    \,    \tau_N \bigl( \Sigma(p_1,p_2,p_3) \bigr)
    \\
    =
    \frac{\mathrm{e}^{\pi \mathrm{i}/4}}{2 \, \sqrt{2 \, P \, N}}
    \,
    \sum_{\substack{
        n= 0 \\
        N \, \nmid \, n
    }}^{2 \, P \, N-1 }
    \mathrm{e}^{- \frac{1}{2 P N} n^2 \pi \mathrm{i} }
    \frac{
      \prod_{j=1}^3
      \left(
        \mathrm{e}^{\frac{  n}{N p_j} \pi \mathrm{i}}
        -
        \mathrm{e}^{- \frac{ n}{N p_j} \pi \mathrm{i} }
      \right)
    }{
      \mathrm{e}^{\frac{n}{N} \pi \mathrm{i}}
      -
      \mathrm{e}^{ - \frac{ n}{N} \pi \mathrm{i}}
    }
  \end{multline}
  where
  \begin{equation}
    \label{phi_definition}
    \phi
    =
    \phi(p_1, p_2, p_3)
    =
    3 - \frac{1}{P}
    +
    12 \, \bigl(
    s(p_2 \, p_3 , p_1) +
    s(p_1 \, p_3 , p_2) +
    s(p_1 \, p_2 , p_3) 
    \bigr)
  \end{equation}
\end{prop}

\begin{proof}
  This was proved in Refs.~\citen{LawreRozan99a,LRozan96f} for
  a general $n$-fibered manifold,
  but we  give a proof here again for 
  completion.

  The Jones polynomial for a link $\mathcal{L}$ depicted in
  Fig.~\ref{fig:Brieskorn} is given by
  \begin{equation*}
    J_{k_0, k_1,k_2,k_3}(\mathcal{L})
    =
    \frac{1}{\sin (\pi/N)} \cdot
    \frac{
      \prod_{j=1}^3
      \sin
      \left( k_0 \, k_j \, \pi/N \right)
    }{
      \sin^2 \left(
        k_0 \, \pi / N
      \right)
    }
  \end{equation*}
  where
  $k_0$ is a color  of an unknotted component whose linking number with other components
  is $1$, and
  $k_j$ (for $j=1,2,3$) denotes a color of a
  component of link $\mathcal{L}$
  which is to be $p_j/q_j$-surgery.
  With this setting we have
  \begin{equation*}
    \sign(\mathbf{L})
    = 
    \sum_{j=1}^3 \sign \left(\frac{q_j}{p_j}\right)
    -1 
  \end{equation*}
  {}From~\eqref{define_invariant}  we get the quantum invariant as
  \begin{multline*}
    \left(
      \mathrm{e}^{\pi \mathrm{i}/N}
      -
      \mathrm{e}^{- \pi \mathrm{i}/N}
    \right) \cdot
    \tau_N \bigl(
    \Sigma(p_1, p_2, p_3)
    \bigr)
    \\
    =
    \mathrm{e}^{
      \frac{\pi \mathrm{i}}{4} \frac{N-2}{N}
      \left(
        3  + \sum_{j=1}^3 \Phi(S U^{(p_j,q_j)})
      \right)
    }
    \sum_{k_0=1}^{N-1}
    \frac{
      \rho(S)_{k_0,1}
    }{
      \left(
        \mathrm{e}^{\frac{\pi \mathrm{i}}{N} k_0}
        -
        \mathrm{e}^{- \frac{\pi \mathrm{i}}{N} k_0}
      \right)^2
    } \,
    \\
    \times
    \prod_{j=1}^3 \sum_{k_j=1}^{N-1}
    \rho(U^{(p_j,q_j)})_{k_j,1} \,
    \left(
      \mathrm{e}^{\frac{\pi \mathrm{i}}{N} k_0 k_j}
      -
      \mathrm{e}^{ -\frac{\pi \mathrm{i}}{N} k_0 k_j}
    \right)
  \end{multline*}
  In this expression we have by definition
  \begin{equation*}
    \rho(S)_{k_0,1}
    =
    \frac{- \mathrm{i}}{\sqrt{2 \, N}} \,
    \left(
      \mathrm{e}^{\frac{\pi \mathrm{i}}{N} k_0}
      -
      \mathrm{e}^{-\frac{\pi \mathrm{i}}{N} k_0}
    \right)
  \end{equation*}
  and for $\ell \in \mathbb{Z}$ we have
  \begin{align*}
    & \sum_{k=1}^{N-1}
    \rho(U^{(p,q)})_{k,1} \,
    \left(
      \mathrm{e}^{\frac{\pi \mathrm{i}}{N} \ell k}
      -
      \mathrm{e}^{ -\frac{\pi \mathrm{i}}{N} \ell k}
    \right)
    \\
    & =
    -\mathrm{i} \,
    \mathrm{e}^{-\frac{\pi \mathrm{i}}{4} \Phi(U^{(p,q)})
      +
      \frac{\pi \mathrm{i} s}{2 N q}}
    \,
    \frac{\sign(q)}{\sqrt{2 \, N \, | q|}} \,
    \sum_{k=1}^{N-1}
    \sum_{
      \substack{
        \gamma \mod 2 N q \\
        \gamma \equiv k \mod 2 N}}
    \mathrm{e}^{\frac{ \pi \mathrm{i} p}{2 N q} \gamma^2}
    \left(
      \mathrm{e}^{\frac{\pi \mathrm{i}}{N q} \gamma}
      -
      \mathrm{e}^{ -\frac{\pi \mathrm{i}}{N q} \gamma}
    \right) \,
    \left(
      \mathrm{e}^{\frac{\pi \mathrm{i}}{N} \ell \gamma}
      -
      \mathrm{e}^{ -\frac{\pi \mathrm{i}}{N} \ell \gamma}
    \right)       
    \\
    & =
    -\mathrm{i} \,
    \mathrm{e}^{-\frac{\pi \mathrm{i}}{4} \Phi(U^{(p,q)})
      +
      \frac{\pi \mathrm{i} s}{2 N q}
    }
    \,
    \frac{\sign(q)}{\sqrt{2 \, N \, | q|}} \,
    \sum_{\gamma \mod 2 N q}
    \mathrm{e}^{\frac{\pi \mathrm{i}}{2 N q} p \gamma^2} \,
    \left(
      \mathrm{e}^{2 \pi \mathrm{i} \frac{q \ell +1 }{2 N q}  \gamma}
      -
      \mathrm{e}^{2 \pi \mathrm{i} \frac{q \ell -1 }{2 N q}  \gamma}
    \right)
    \\
    & =
    -\frac{\sign(p)}{\sqrt{|p|}} \,
    \mathrm{e}^{
      -\frac{\pi \mathrm{i}}{4} 
      \Phi(S U^{(p,q)})
      +
      \frac{\pi \mathrm{i} s}{2 N q}
    }
    \sum_{n \mod p}
    \left(
      \mathrm{e}^{-\frac{\pi \mathrm{i}}{2 p q N}
        (2 N q n + q \ell+1)^2
      }
      -
      \mathrm{e}^{-\frac{\pi \mathrm{i}}{2 p q N}
        (2 N q n + q \ell-1)^2
      }
    \right)
  \end{align*}
  Here in the first equality we have used
  $\gamma=k \pmod{ 2 \, N}$, and applied a symmetry under
  $\gamma \to 2 \, N \, q- \gamma$ in the second equality.
  In the last equality, we have used
  an identity
  $\mathrm{e}^{\frac{\pi \mathrm{i}}{2} (1-\sign(p))}
  =
  \sign(p)
  $,
  and the Gauss sum reciprocity formula~\cite{LCJeff92a}
  \begin{equation}
    \label{Gauss_sum_formula}
    \sum_{n \mod N}
    \mathrm{e}^{\frac{\pi \mathrm{i}}{N} M n^2 + 2 \pi \mathrm{i} k n}
    =
    \sqrt{
      \left|
        \frac{N}{M}
      \right|
    } \,
    \mathrm{e}^{\frac{\pi \mathrm{i}}{4} \sign(N M)}
    \sum_{n \mod M}
    \mathrm{e}^{
      -\frac{\pi \mathrm{i}}{M}
      N (n+k)^2
    }
  \end{equation}
  where $N, M\in \mathbb{Z}$
  with
  $N \, k \in \mathbb{Z}$ and
  $N \, M$ being even.

  A combination of  these results reduces to
  \begin{multline*}
    \left(
      \mathrm{e}^{\pi \mathrm{i}/N}
      -
      \mathrm{e}^{- \pi \mathrm{i}/N}
    \right) \cdot
    \tau_N \bigl(
    \Sigma(p_1, p_2, p_3)
    \bigr)
    \\
    =
    \frac{\mathrm{i} \, \sign(P)}{
      \sqrt{2 \, |P| \, N}}
    \,
    \mathrm{e}^{
      \frac{3}{4} \pi \mathrm{i} 
      +\frac{\pi \mathrm{i}}{2 N} \sum_j \frac{1}{p_j q_j}
      -
      \frac{\pi \mathrm{i}}{2 N} \phi
    }
    \sum_{k_0=1}^{N-1}
    \sum_{n_j \mod p_j}
    \frac{1}{
      \mathrm{e}^{\frac{\pi \mathrm{i}}{N} k_0}
      -
      \mathrm{e}^{- \frac{\pi \mathrm{i}}{N} k_0}
    }
    \,
    \\
    \times
    \prod_{j=1}^3
    \left(
      \mathrm{e}^{-\frac{\pi \mathrm{i}}{2 N p_j q_j}
        (2 N q_j n_j + k_0 q_j +1)^2
      }
      -
      \mathrm{e}^{-\frac{\pi \mathrm{i}}{2 N p_j q_j}
        (2 N q_j n_j + k_0 q_j -1)^2
      }
    \right)
  \end{multline*}
  The summand is invariant under
  (i)
  $k_0 \to k_0 + 2 \, N$
  and
  $n_j\to n_j -1$,
  (ii)
  $n_j  \to n_j + p_j$.
  Using this symmetry and
  recalling that $p_j$ are pairwise coprime integers,
  the sum,
  $\sum_{k_0=1}^{N-1}
  \sum_{n_j \mod{ p_j}}$,
  is transformed into
  a sum,
  $\sum_{
    \substack{
      k_0 = a + 2 N n \\
      1 \leq a \leq N-1 \\
      0 \leq n \leq P-1
    }}
  $, with setting all $n_j=0$.
  As a result, we find
  \begin{multline*}
    \left(
      \mathrm{e}^{\pi \mathrm{i}/N}
      -
      \mathrm{e}^{- \pi \mathrm{i}/N}
    \right) \cdot
    \tau_N \bigl(
    \Sigma(p_1, p_2, p_3)
    \bigr)
    \\
    =
    -\frac{\mathrm{i}}{
      \sqrt{2 \, P \, N}}
    \,
    \mathrm{e}^{
      \frac{3}{4} \pi \mathrm{i} 
      -
      \frac{\pi \mathrm{i}}{2 N} \phi
    }
    \,
    \sum_{
      \substack{
        k_0 = a + 2 N n \\
        1 \leq a \leq N-1 \\
        0 \leq n \leq P-1
      }}
    \mathrm{e}^{-\frac{\pi \mathrm{i}}{2 N P} k_0^{~2}}
    \,
    \frac{
    \prod_{j=1}^3
    \left(
      \mathrm{e}^{\frac{\pi \mathrm{i}}{N p_j} k_0}
      -
      \mathrm{e}^{-\frac{\pi \mathrm{i}}{N p_j} k_0}
    \right)
    }{
      \mathrm{e}^{\frac{\pi \mathrm{i}}{N} k_0}
      -
      \mathrm{e}^{- \frac{\pi \mathrm{i}}{N} k_0}
    }
  \end{multline*}
  Setting $k_0\to 2 \, P \, N- k_0$, we obtain a statement of the proposition.
\end{proof}

\section{Modular Forms}
\label{sec:modular}

We define the odd periodic function
$\chi_{2 P}^{(\ell_1, \ell_2 ,  \ell_3)}(n)$
with modulus $2 \, P$
by
\begin{equation}
  \label{define_chi}
  \chi_{2 P}^{(\ell_1, \ell_2 ,  \ell_3)}(n)
  =
  \begin{cases}
    1 &
    \text{for
      $\displaystyle
      n=
      P \,
      \Bigl(
        1 + \sum_{j=1}^3 \varepsilon_j \, \frac{\ell_j}{p_j}
      \Bigr)
      \mod 2 \, P
      $
      where
      $\varepsilon_1 \varepsilon_2 \varepsilon_3=-1$
    }
    \\[2mm]
    -1 &
    \text{for
      $\displaystyle
      n=
      P \, 
      \Bigl(
        1 + \sum_{j=1}^3 \varepsilon_j \, \frac{\ell_j}{p_j}
      \Bigr)
      \mod 2 \, P
      $
      where
      $\varepsilon_1 \varepsilon_2 \varepsilon_3= 1$
    }
    \\[2mm]
    0 &
    \text{others}
  \end{cases}
\end{equation}
Here
$P=p_1 \, p_2 \, p_3$ with pairwise coprime positive integers $p_j$,
and
we mean $\varepsilon_j = \pm 1$.
Integers  $\ell_j$ are 
\begin{equation}
  \label{define_ell}
  1 \leq \ell_j \leq p_j - 1
\end{equation}
There exists a symmetry  of the periodic function
\begin{align}
  \chi_{2 P}^{(\ell_1, \ell_2, \ell_3)}(n)
  & =
  \chi_{2 P}^{(p_1-\ell_1, p_2 - \ell_2, \ell_3)}(n)
  \nonumber
  \\
  & =
  \chi_{2 P}^{(p_1-\ell_1, \ell_2, p_3 - \ell_3)}(n)
  =
  \chi_{2 P}^{(\ell_1, p_2 - \ell_2, p_3 - \ell_3)}(n)
  \label{symmetry_chi}
\end{align}

With this periodic function, we define the function
$  \Phi_{\boldsymbol{p}}^{(\ell_1, \ell_2, \ell_3)}(\tau)$
for $\tau$  in the upper half plane,
$\tau \in \mathbb{H}$,  by
\begin{equation}
  \label{modular_Phi}
  \Phi_{\boldsymbol{p}}^{(\ell_1, \ell_2, \ell_3)}(\tau)
  =
  \frac{1}{2}
  \sum_{n \in \mathbb{Z}}
  n \,
  \chi_{2 P}^{(\ell_1, \ell_2, \ell_3)}(n)
  \,
  q^{\frac{n^2}{4 P}}
\end{equation}
where  as usual 
\begin{equation*}
  q=\exp(2 \, \pi \, \mathrm{i} \, \tau)
\end{equation*}
Eq.~\eqref{symmetry_chi}
makes the number of the independent
functions $\Phi_{\boldsymbol{p}}^{(\ell_1, \ell_2, \ell_3)}(\tau)$
to be
\begin{equation}
  D
  =
  D(p_1, p_2, p_3)
  =\frac{1}{4} \,
  (p_1 - 1) \,   (p_2 - 1) \,   (p_3- 1) 
\end{equation}

\begin{prop}
  The function $\Phi_{\boldsymbol{p}}^{(\ell_1, \ell_2, \ell_3)}(\tau)$ is a
  modular form with weight $3/2$.
  Namely under the ${S}$- and ${T}$-transformations~\eqref{PSL_ST}
  we have
  \begin{align}
    \label{S_transformation}
    \Phi_{\boldsymbol{p}}^{(\ell_1, \ell_2, \ell_3)}(\tau)
    & =
    \left(
      \frac{\mathrm{i}}{\tau}
    \right)^{3/2} \,
    \sum_{\ell_1^\prime, \ell_2^\prime, \ell_3^\prime}
    \mathbf{S}_{\ell_1, \ell_2, \ell_3}^{\ell_1^\prime, \ell_2^\prime,
      \ell_3^\prime} \,
    \Phi_{\boldsymbol{p}}^{(\ell_1^\prime, \ell_2^\prime,
      \ell_3^\prime)}(-1/\tau)
    \\[2mm]
    \Phi_{\boldsymbol{p}}^{(\ell_1, \ell_2, \ell_3)}(\tau+1)
    & =
    \mathbf{T}^{\ell_1, \ell_2, \ell_3}
    \Phi_{\boldsymbol{p}}^{(\ell_1, \ell_2, \ell_3)}(\tau)
  \end{align}
  where the sum runs over $D(p_1,p_2,p_3)$ distinct triples.
  A $D\times D$ matrix $\mathbf{S}$ and
  diagonal matrix $\mathbf{T}$  are respectively given by
  \begin{align}
    \mathbf{S}_{\ell_1, \ell_2, \ell_3}^{\ell_1^\prime, \ell_2^\prime,
      \ell_3^\prime} 
    & =
    \sqrt{\frac{32}{P}} \,
    (-1)^{1+ P  + P  \sum_{j=1}^3 \frac{\ell_j + \ell_j^\prime}{p_j}
      + (\boldsymbol{\ell} \times \boldsymbol{\ell^\prime}) \cdot
      \boldsymbol{p}
    } \,
    \prod_{j=1}^3
    \sin \left(
      P \, \frac{\ell_j \, \ell_j^\prime}{p_j^{~2}}  \, \pi
    \right)
    \\[2mm]
    \mathbf{T}^{\ell_1, \ell_2, \ell_3}
    & =
    \exp \left(
      \frac{\pi \, \mathrm{i} }{2} \, P \,
      \Bigl(
      1+ \sum_{j=1}^3 \frac{\ell_j}{p_j}
      \Bigr)^2
    \right)
  \end{align}
\end{prop}

We omit a proof as it is tedious but straightforward.
We only need the Poisson summation formula
\begin{equation}
  \sum_{n \in \mathbb{Z}} f(n)
  =
  \sum_{n \in \mathbb{Z}}
  \int_{-\infty}^\infty
  \mathrm{e}^{-2 \pi \mathrm{i} t n} \,
  f(t) \,
  \mathrm{d} t
\end{equation}

\section{The Eichler Integral and the Chern--Simons Invariant}
\label{sec:Eichler}

The Eichler integral was originally defined 
as a $k-1$-fold integration of  a modular form with integral
weight $k\in \mathbb{Z}_{\geq 2}$
(see, \emph{e.g.}, Ref.~\citen{SLang76Book}).
Following Refs.~\citen{LawrZagi99a,DZagie01a}
(see also Refs.~\citen{KHikami02c,KHikami03a,KHikami03c}), 
we define the Eichler integral of the modular form
$\Phi_{\boldsymbol{p}}^{(\ell_1,\ell_2,\ell_3)}(\tau)$ with \emph{half-integral}
weight $3/2$ by
\begin{equation}
  \widetilde{\Phi}_{\boldsymbol{p}}^{(\ell_1, \ell_2, \ell_3)}(\tau)
  =
  \sum_{n=0}^\infty
  \chi_{2 P}^{(\ell_1 , \ell_2, \ell_3)}(n) \,
  q^{\frac{n^2}{4 P}}
\end{equation}
We should remark that there are 
$D(p_1,p_2,p_3)$ independent Eichler integrals due to the
symmetry~\eqref{symmetry_chi}.

\begin{prop}
  \label{prop:Eichler_limit}
  The function
  $
  \widetilde{\Phi}_{\boldsymbol{p}}^{(\ell_1, \ell_2, \ell_3)}(\tau)
  $
  has a limiting value in $\tau \to 1/N$ for
   $N\in \mathbb{Z}$ as
  \begin{equation}
    \label{Eichler_fraction}
    \widetilde{\Phi}_{\boldsymbol{p}}^{(\ell_1, \ell_2, \ell_3)}(1/N)
    =
    \sum_{n=0}^{P N}
    \chi_{2 P}^{(\ell_1 , \ell_2, \ell_3)}(n) \,
    \left(
      1 - \frac{n}{P \, N}
    \right) \,
    \mathrm{e}^{\frac{1}{2 P N} n^2 \pi \mathrm{i}}
  \end{equation}
  We also have
  \begin{align}
    \label{Eichler_integer}
    \widetilde{\Phi}_{\boldsymbol{p}}^{(\ell_1, \ell_2, \ell_3)}(N)
    & =
    - \frac{1}{2 \, P} \,
    \left(
      \sum_{n=1}^{2 P}
      n \,
      \chi_{2 P}^{(\ell_1 , \ell_2, \ell_3)}(n) \,
    \right) \,
    \mathrm{e}^{\frac{\pi \mathrm{i}}{2} P N
      \left(
        1+\sum_j \frac{\ell_j}{p_j}
      \right)^2
    }
    \\
    & =
    - \frac{1}{2 \, P} \,
    \left(
      \sum_{n=1}^{2 P}
      n \,
      \chi_{2 P}^{(\ell_1 , \ell_2, \ell_3)}(n) \,
    \right) \,
    \left(
      \mathbf{T}^{\ell_1 , \ell_2, \ell_3}
    \right)^N
    \nonumber
  \end{align}
\end{prop}

To prove this proposition, we use the following formula for asymptotic expansions
(see Refs.~\citen{LawrZagi99a,DZagie01a});
\begin{prop}
  \label{prop:L_function}
  Let $C_f(n)$ be a periodic function with mean value $0$ and modulus $f$.
  Then we have an  asymptotic expansion  as $t \searrow 0$;
  \begin{gather}
    \sum_{n=1}^\infty C_f(n) \, \mathrm{e}^{-n   t}
    \simeq
    \sum_{k=0}^\infty
    L(- k, C_f) \,
    \frac{(-t)^k}{k!} 
    \\[2mm]
    \sum_{n=1}^\infty C_f(n) \, \mathrm{e}^{-n^2  t}
    \simeq
    \sum_{k=0}^\infty
    L(-2 \, k, C_f) \,
    \frac{(-t)^k}{k!} 
  \end{gather}
  Here
  $L(k,C_f)$ is the Dirichlet $L$-function associated with $C_f(n)$,
  and is given by
  \begin{equation*}
    L(-k, C_f)
    =
    - \frac{f^k}{k+1} \,
    \sum_{n=1}^f
    C_f(n) \, B_{k+1} \left( \frac{\ n \ }{f} \right) 
  \end{equation*}
  where
  $B_n(x)$ is the $n$-th Bernoulli polynomial defined from
  \begin{equation*}
    \frac{
      t \, \mathrm{e}^{x t}
    }{
      \mathrm{e}^t -1
    }
    =
    \sum_{n=0}^\infty
    \frac{B_n(x)}{n!} \,
    t^n
  \end{equation*}
\end{prop}

See, \emph{e.g.}, Ref.~\citen{LawrZagi99a} for a proof.

\begin{proof}[Proof of Proposition~\ref{prop:Eichler_limit}]
  We assume $M$ and $N$ are coprime integers, and $N>0$.
  By definition, we have
  \begin{equation*}
    \widetilde{\Phi}_{\boldsymbol{p}}^{(\ell_1,\ell_2,\ell_3)}
    \left(
    \frac{M}{N} + \mathrm{i} \, \frac{y}{2 \, \pi}
    \right)
    =
    \sum_{n=0}^\infty C_{2 P N}^{(\ell_1,\ell_2,\ell_3)}(n) \,
    \mathrm{e}^{- \frac{y}{4 P} n^2}
  \end{equation*}
  where $y>0$ and
  \begin{equation*}
    C_{2 P N}^{(\ell_1,\ell_2,\ell_3)}(n) 
    =
    \chi_{2 P}^{(\ell_1,\ell_2,\ell_3)}(n) \,
    \mathrm{e}^{\frac{M}{2 P N} n^2 \pi \mathrm{i}}
  \end{equation*}
  We see that
  $ C_{2 P N}^{(\ell_1,\ell_2,\ell_3)}(n+2 \, P \, N) 
  =  C_{2 P N}^{(\ell_1,\ell_2,\ell_3)}(n) $,
  and
  $ C_{2 P N}^{(\ell_1,\ell_2,\ell_3)}(2 \, P \, N -n) 
  = - C_{2 P N}^{(\ell_1,\ell_2,\ell_3)}(n) $.
  Then we can apply Prop.~\ref{prop:L_function}, and
  we have an asymptotic expansion in $y\searrow 0$ as
  \begin{equation*}
    \widetilde{\Phi}_{\boldsymbol{p}}^{(\ell_1,\ell_2,\ell_3)}
    \left(
    \frac{M}{N} + \mathrm{i} \, \frac{y}{2 \, \pi}
    \right)
    \simeq
    \sum_{k=0}^\infty
    \frac{L(-2 \, k , C_{2 P N}^{(\ell_1,\ell_2,\ell_3)})}{k!} \,
    \left(
      - \frac{y}{4 P}
    \right)^k
  \end{equation*}
  which gives a limiting value as
  \begin{equation}
    \label{Eichler_and_L}
    \widetilde{\Phi}_{\boldsymbol{p}}^{(\ell_1,\ell_2,\ell_3)}
    (M/N)
    =
    L(0 , C_{2 P N}^{(\ell_1,\ell_2,\ell_3)})
  \end{equation}
  Using a fact that 
  $\chi_{2P}^{(\ell_1,\ell_2,\ell_3)}(2 \, P - n)
  =
  -  \chi_{2P}^{(\ell_1,\ell_2,\ell_3)}(n)$
  and that
  an explicit form of the Bernoulli polynomial is
  $B_1(x)=x-\frac{1}{2}$,
  we get
  \begin{equation*}
    \widetilde{\Phi}_{\boldsymbol{p}}^{(\ell_1,\ell_2,\ell_3)}
    (M/N)
    =
    \sum_{n=0}^{P N}
    \chi_{2 P}^{(\ell_1,\ell_2,\ell_3)}(n) \,
    \left(
      1- \frac{n}{P \, N}
    \right) \,
    \mathrm{e}^{
      \frac{M}{2 P N} n^2 \pi \mathrm{i}
    }
  \end{equation*}
  Eq.~\eqref{Eichler_fraction} directly follows from this formula.
  Eq.~\eqref{Eichler_integer} can also be given from the above formula when we recall
  $\sum_{n=1}^{2 P}       \chi_{2 P}^{(\ell_1 , \ell_2, \ell_3)}(n) 
  =0$.
\end{proof}

We can see that,
though we have  $D(p_1, p_2, p_3)$ independent Eichler integrals,
the  limiting value
$\widetilde{\Phi}_{\boldsymbol{p}}^{(\ell_1, \ell_2, \ell_3)}(N)$
at  $N\in \mathbb{Z}$
computed in eq.~\eqref{Eichler_integer}
becomes identically  zero for some triples $(\ell_1,\ell_2,\ell_3)$.

\begin{prop}
  Let  $\gamma(p_1,p_2,p_3)$  be
  the number of
  independent Eichler integrals
  such that
  $\widetilde{\Phi}_{\boldsymbol{p}}^{(\ell_1,\ell_2,\ell_3)}(N) \not\equiv 0$
  for $N\in\mathbb{Z}$, namely
  \begin{equation}
    \label{zero_condition}
    \sum_{n=1}^{2 P} n \, 
    \chi_{2 P}^{(\ell_1,\ell_2,\ell_3)}(n)
    \neq 0
  \end{equation}
  We then have
  \begin{multline}
    \label{value_gamma}
    \gamma(p_1, p_2, p_3)
    =
    s(p_1 \, p_2, p_3)+    s(p_2 \, p_3, p_1)+    s(p_1 \, p_3, p_2)
    \\
    +
    \frac{P}{12} \,
    \left(
      1-
      \frac{1}{p_1^{~2}}
      -   \frac{1}{p_2^{~2}}
      -  \frac{1}{p_3^{~2}}
    \right)
    -\frac{1}{12  \, P}
    +
    \frac{1}{4}
  \end{multline}
  where $s(b,a)$ is the Dedekind sum defined in eq.~\eqref{Dedekind_sum}.
\end{prop}

\begin{proof}
  For a sake of our brevity we set
  $A_k=\frac{\ell_i}{p_i}+\frac{\ell_j}{p_j}-\frac{\ell_k}{p_k}$
  for $i\neq j \neq k \neq i$ and $i,j,k\in \{1,2,3\}$.
  As we have $0 < \frac{\ell_k}{p_k}<1$ by
  definition,
  we have
  $0 < \sum_j  \frac{\ell_j}{p_j} < 3$
  and
  $-1 < A_k <2$.

  When $ 0<  \sum_j \frac{\ell_j}{p_j} < 1$,
  we have
  $
  0< A_k  +1 < 2 \left( 1-\frac{\ell_k}{p_k} \right)<2$.
  Then in a domain   $0<n<2 \, P$,
  the periodic function
  $\chi_{2 P}^{(\ell_1,\ell_2,\ell_3)}(n)$
  defined in eq.~\eqref{define_chi}
  takes a value $1$ when
  $n= P \, \left(1-\sum_j \frac{\ell_j}{p_j} \right),
  P (1 + A_k)$,
  while it is
  $-1$
  when
  $n= P \, \left( 1+\sum_j \frac{\ell_j}{p_j} \right),
  P \, \left( 1 - A_k \right)$.
  Then  we find
  $\sum_{n=1}^{2 P} n \chi_{2 P}^{(\ell_1,\ell_2,\ell_3)}(n)=0$,
  and it is inconsistent with eq.~\eqref{zero_condition}.

  We thus have a condition
  $1 < \sum_j  \frac{\ell_j}{p_j} < 3$ to fulfill
  eq.~\eqref{zero_condition},
  because it is impossible to have
  $\sum_j \frac{\ell_j}{p_j} \in \mathbb{Z}$.
  Under  this condition there are two possibilities for a condition of
  $A_k$;
  (i)
  $-1 < A_k <1$ for all $k$,
  or
  (ii)
  $-1 < A_k <1$ for two $k$'s and
  $1<A_i<2$ for another $i$.
  By the same computation we can check
  \begin{equation}
    \label{non_zero_integer}
    \sum_{n=1}^{2 P} n \, \chi_{2 P}^{(\ell_1,\ell_2,\ell_3)}(n)=4 \, P
  \end{equation}
  for the former case, while
  we have
  $\sum_{n=1}^{2 P} n \chi_{2 P}^{(\ell_1,\ell_2,\ell_3)}(n)=0$
  for the latter.
  To conclude,  a condition~\eqref{zero_condition} is fulfilled when
  the triple of integers satisfies
  \begin{equation}
    \label{ell_condition}
    \begin{aligned}
      & 1<\frac{\ell_1}{p_1}+ \frac{\ell_2}{p_2} + \frac{\ell_3}{p_3}<3
      & &
      -1<\frac{\ell_1}{p_1}+ \frac{\ell_2}{p_2} - \frac{\ell_3}{p_3}<1
      \\[2mm]
      &
      -1<\frac{\ell_1}{p_1} - \frac{\ell_2}{p_2} + \frac{\ell_3}{p_3}<1
      & &
      -1<- \frac{\ell_1}{p_1}+ \frac{\ell_2}{p_2} + \frac{\ell_3}{p_3}<1
    \end{aligned}
  \end{equation}
  This constraint is depicted as the number of the integral lattice
  points
  of an interior of the tetrahedron
  (see Fig.~\ref{fig:tetrahedron}).

  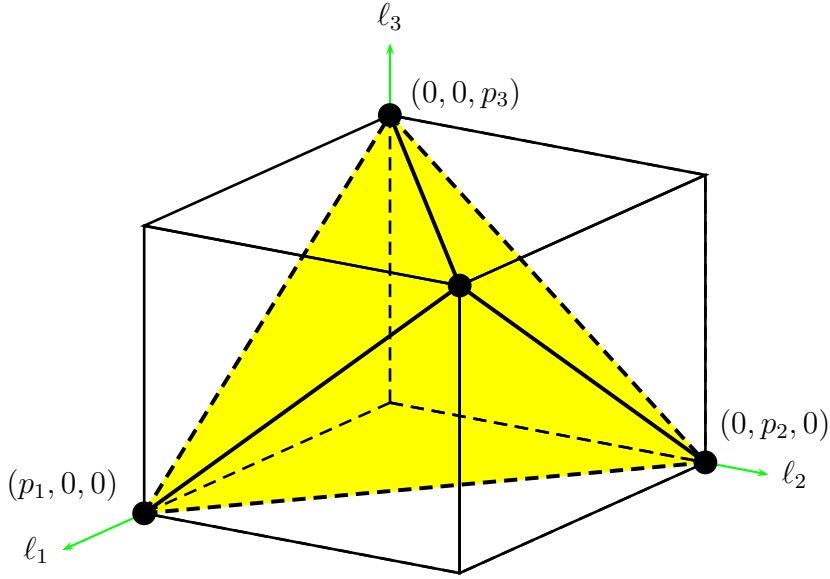
\begin{figure}[htbp]
    \centering
      \begin{pspicture}(-4,-2.2)(4,5)
        \psset{
          Alpha=57,Beta=17,xMin=0,xMax=8,yMin=0,yMax=6,zMin=0,zMax=5,
        }
        \pstThreeDCoor[
        linecolor=green,
        linewidth=0.8pt,
        nameX=\text{$\ell_1$},
        nameY=\text{$\ell_2$},
        nameZ=\text{$\ell_3$},
        ]
%
        \pstThreeDTriangle[fillstyle=solid,fillcolor=yellow,linecolor=yellow](6,5,4)(6,0,0)(0,5,0)
        \pstThreeDTriangle[fillstyle=solid,fillcolor=yellow,linecolor=yellow](6,5,4)(6,0,0)(0,0,4)
        \pstThreeDTriangle[fillstyle=solid,fillcolor=yellow,linecolor=yellow](6,5,4)(0,0,4)(0,5,0)
        \pstThreeDBox[linewidth=1.0pt](0,0,0)(6,0,0)(0,5,0)(0,0,4)
        \pstThreeDTriangle[%
        linecolor=black,linestyle=dashed,
        linewidth=1.5pt](6,0,0)(0,5,0)(0,0,4)
        \pstThreeDLine[linewidth=1.5pt](6,5,4)(0,0,4)
        \pstThreeDLine[linewidth=1.5pt](6,5,4)(0,5,0)
        \pstThreeDLine[linewidth=1.5pt](6,5,4)(6,0,0)
        \psset{dotstyle=*,dotscale=3}
        \pstThreeDDot(6,0,0)
        \pstThreeDDot(0,5,0)
        \pstThreeDDot(0,0,4)
        \pstThreeDDot(6,5,4)
        \pstThreeDPut(8,0,0.9){$(p_1,0,0)$}
        \pstThreeDPut(0,6.1,0.7){$(0,p_2,0)$}
        \pstThreeDPut(0,1.2,4.5){$(0,0,p_3)$}
      \end{pspicture}
    \caption{The number of the integral lattice points in the
      tetrahedron     coincides with $\gamma(p_1,p_2,p_3)$.
    }
    \label{fig:tetrahedron}
  \end{figure}

  Let
  $N_3(p_1,p_2,p_3)$ be the number of the integral lattice points
  $(\ell_1, \ell_2, \ell_3)$  such that $0<\ell_k<p_k$ and
  $0< \sum_j \frac{\ell_j}{p_j} <1$.
  Then by the symmetry of the triple,
  \begin{equation}
    \label{ell_symmetry}
    \begin{aligned}
      & (\ell_1,\ell_2,\ell_3) 
      & & (\ell_1, p_2 - \ell_2, p_3  -\ell_3) 
      \\
      & (p_1 -\ell_1,  \ell_2, p_3 - \ell_3) 
      & & (p_1 - \ell_1, p_2 - \ell_2, \ell_3) 
    \end{aligned}
  \end{equation}
  we have
  \begin{equation*}
    \gamma(p_1, p_2, p_3 )
    =
    D(p_1, p_2 , p_3) - N_3 (p_1 , p_2, p_3)
  \end{equation*}
  It is well known that
  $N_3(p_1, p_2, p_3)$ was computed by Mordell~\cite{Morde51} (see also
  Ref.~\citen{RademGross72}),
  and by substituting this result, we get
  eq.~\eqref{value_gamma}.
\end{proof}

The Casson invariant $\lambda_C(\mathcal{M})$
for the Brieskorn homology spheres
 $\mathcal{M}= \Sigma(p_1, p_2, p_3)$
is given by~\cite{FintuStern90a,FukuMatsSaka90}
(also see, \emph{e.g.}, Ref.~\citen{NSave99Book})
\begin{multline}
  \label{Casson_sphere}
  \lambda_C\bigl(\Sigma(p_1, p_2, p_3) \bigr)
  =
  -\frac{1}{2}
  \left(
    s(p_1 \, p_2, p_3)+    s(p_2 \, p_3, p_1)+    s(p_1 \, p_3, p_2)
  \right)
    \\
    -
    \frac{P}{24} \,
    \left(
      1-
      \frac{1}{p_1^{~2}}
      -   \frac{1}{p_2^{~2}}
      -  \frac{1}{p_3^{~2}}
    \right)
    + \frac{1}{24  \, P}
    -
    \frac{1}{8}
\end{multline}
With this result, we have the following theorem.

\begin{theorem}
  The Casson invariant is a minus half of the number of the Eichler
  integrals which 
  do not vanish at $\tau\to N\in\mathbb{Z}$;
  \begin{equation}
    \label{gamma_Casson}
    \lambda_C \bigl(\Sigma(p_1, p_2, p_3) \bigr)
    =
    -\frac{1}{2} \,
    \gamma(p_1, p_2, p_3)
  \end{equation}
\end{theorem}

The fundamental group of the Brieskorn homology spheres has the
presentation~\eqref{presentation}, and the Casson
invariant is related to the representation space of the fundamental
group~\cite{FintuStern90a}.
That is to say,
we  see that
$\gamma(p_1, p_2, p_3)$ coincides with the cardinality of the
SU(2) representation space of
$\pi_1\bigl( \Sigma(p_1, p_2, p_3) \bigr)$,
and that the triple $(\ell_1, \ell_2, \ell_3)$ has a one-to-one
correspondence with the irreducible representation.
Explicitly when we have
an irreducible representation $\alpha$ of
$\pi_1\bigl( \Sigma(p_1, p_2, p_3) \bigr)$,
the conjugacy class of $\alpha(x_k)$ is
\begin{equation}
  \label{conjugacy}
  \begin{pmatrix}
    \mathrm{e}^{\frac{p_k - \ell_k}{p_k} \pi \mathrm{i}} &
    \\
    & \mathrm{e}^{- \frac{p_k - \ell_k}{p_k} \pi \mathrm{i}} 
  \end{pmatrix}
\end{equation}
where the triple of integers $(\ell_1, \ell_2, \ell_3)$
satisfies  a  condition~\eqref{ell_condition}.

Using this correspondence with the irreducible representation of the fundamental group,
we can read off the Chern--Simons invariant
of the Brieskorn homology spheres~\cite{FintuStern90a}
(see also Ref.~\citen{KirkKlas90a});
\begin{equation}
  \label{CS_B}
  \CS(A)
  =
  - \frac{P}{4} \, 
  \left(
    1+ \sum_{j=1}^3 \frac{\ell_j}{p_j}
  \right)^2
  \mod 1
\end{equation}

\begin{theorem}
  The Chern--Simons invariant~\eqref{CS_B} for the Brieskorn homology
  spheres
  $\Sigma(p_1,p_2,p_3)$ is related to a limit value of the Eichler
  integral which can be regarded as the $\mathbf{T}$-matrix of the modular group;
  for $N\in\mathbb{Z}$ we have
  \begin{align}
    \widetilde{\Phi}_{\boldsymbol{p}}^{(\ell_1,\ell_2,\ell_3)}
    (N)
    & =
    -2 \,
    \left(
      \mathbf{T}^{\ell_1 , \ell_2, \ell_3}
    \right)^N
    \nonumber
    \\
    & =
    - 2 \,
    \mathrm{e}^{ - 2 \pi \mathrm{i} \CS(A) N}
    \label{integer_Chern}
  \end{align}
  where the triple of integers $(\ell_1,\ell_2,\ell_3)$
  satisfies eq.~\eqref{ell_condition}.
  The  triple $(\ell_1,\ell_2,\ell_3)$ also gives a
  $SU(2)$ representation
  $\alpha$ of the
  fundamental group~\eqref{presentation},
  where
  the  conjugacy class of  $\alpha(x_k)$ is 
  as in eq.~\eqref{conjugacy}.
\end{theorem}

We should point out that,
with $s$ and $t$ being  coprime positive integers.
the number,
$(s-1) \, (t-1)/2$,
 of integral lattice points
in the 2-dimensional space
has appeared as
the number of the irreducible  highest weight representation of the
Virasoro algebra of the minimal model
$\mathcal{M}(s,t)$ 
in the conformal
field theory~\cite{BPZ84a}.
The  character of the minimal model $\mathcal{M}(s,t)$
is modular with weight $1/2$, and it was
shown~\cite{KHikami03c,KHikami03b} that the Eichler integral thereof
is a specific value of the colored Jones polynomial for the torus knot
$\mathcal{T}_{s,t}$.
Also
we may say that
a one-dimensional analogue of the number of the integral lattice
points
is realized in a torus link;
the colored Jones polynomial for the torus link
$\mathcal{T}_{2,2  N}$
coincides with the Eichler integral of the $\widehat{su}(2)_{N-2}$
character~\cite{KHikami03a},
which is a ($N-1$)-dimensional representation~\eqref{affine_su}
of the modular group  with weight $3/2$.
In this sense, the Brieskorn homology spheres may be regarded as a
generalization of the torus knot and link
from the viewpoint of the modular form.
This may indicate the fact that the Brieskorn homology sphere
$\Sigma(p_1,p_2,p_3)$
is homeomorphic to the 
$p_3$-fold cyclic branched covering of $S^3$ branched along a torus knot
$\mathcal{T}_{p_1,p_2}$~\cite{JMiln75a},
and that
the manifold
$\Sigma(p, q, p \, q \,n \pm 1)$
can be alternatively
constructed  by $\pm 1/n$-surgery of the torus knot $\mathcal{T}_{p,q}$.

For our purpose to give an exact asymptotic behavior of the WRT invariant,
we shall give an asymptotic expansion of the Eichler integral
at $\tau\to 1/N$.
The Eichler integral is no longer modular, but it has a \emph{nearly}
modular property
when $\tau\in \mathbb{Q}$
as was studied in Ref.~\citen{LawrZagi99a}.
\begin{prop}
  The Eichler integral
  $\widetilde{\Phi}_{\boldsymbol{p}}^{(\ell_1,\ell_2,\ell_3)}(\tau)$
  in $\tau \to 1/N$ for $N\in\mathbb{Z}_{>0}$
  fulfills a nearly modular property.
  Namely, under the $S$-transformation,
  we have an asymptotic expansion in $N\to\infty$ as follows;
  \begin{multline}
    \label{asymptotics_Eichler}
    \widetilde{\Phi}_{\boldsymbol{p}}^{(\ell_1, \ell_2, \ell_3)}(1/N)
    \simeq
    -
    \sqrt{\frac{N}{ \mathrm{i}}}
    \sum_{\ell_1^\prime, \ell_2^\prime, \ell_3^\prime}
    \mathbf{S}_{\ell_1, \ell_2, \ell_3}^{\ell_1^\prime, \ell_2^\prime,
      \ell_3^\prime} \,
    \widetilde{\Phi}_{\boldsymbol{p}}^{(\ell_1^\prime, \ell_2^\prime,
      \ell_3^\prime)}(-N)
    \\
    +
    \sum_{k=0}^\infty
    \frac{
      L(-2 \, k , \chi_{2 P}^{(\ell_1, \ell_2, \ell_3)})
    }{
      k!
    }
    \,
    \left(
      \frac{\pi \, \mathrm{i}}{
        2 \, P \, N
      }
    \right)^k
  \end{multline}
  Here the $L$-function is given by
  \begin{equation}
    \label{L_and_Bernoulli}
    L(-2 \, k , \chi_{2 P}^{(\ell_1, \ell_2, \ell_3)})
    =
    - \frac{(2 \, P)^{2 k}}{2 \, k +1}
    \sum_{j=1}^{2 P}
    \chi_{2 P}^{(\ell_1 , \ell_2 ,\ell_3)}(j) \,
    B_{2 k +1}
    \left( \frac{j}{2 \, P} \right)
  \end{equation}
\end{prop}

\begin{proof}
  A proof is essentially the same with one given in Ref.~\citen{LawrZagi99a}
  (see also Refs.~\citen{KHikami02c,KHikami03a,KHikami03c}).
  We use
  \begin{equation}
    \label{define_Phi_hat}
    \widehat{\Phi}_{\boldsymbol{p}}^{(\ell_1,\ell_2,\ell_3)}(z)
    =
    \frac{1}{\sqrt{2 \, P \, \mathrm{i}}} \,
    \int_{z^*}^\infty
    \frac{
      \Phi_{\boldsymbol{p}}^{(\ell_1,\ell_2,\ell_3)}(\tau)
    }{
      \sqrt{\tau - z}
    }
    \,
    \mathrm{d} \tau
  \end{equation}
  which is defined for $z$ in the lower half plane,
  $z \in \mathbb{H}^-$.
  We mean that  $z^*$ denotes a  complex conjugate,
  and we do not have a singularity in integral.
  Applying the modular ${S}$-transformation~\eqref{S_transformation},
  we see that the function
  $    \widehat{\Phi}_{\boldsymbol{p}}^{(\ell_1,\ell_2,\ell_3)}(z)$
  has a nearly modular property;
  \begin{equation}
    \label{modular_Phi_hat}
    \frac{1}{\sqrt{\mathrm{i} \, z}} \,
    \sum_{\ell_1^\prime, \ell_2^\prime, \ell_3^\prime}
    \mathbf{S}_{\ell_1, \ell_2, \ell_3}^{\ell_1^\prime, \ell_2^\prime,
      \ell_3^\prime} \,
    \widehat{\Phi}_{\boldsymbol{p}}^{(\ell_1^\prime, \ell_2^\prime,
      \ell_3^\prime)}(-1/z)
    +
    \widehat{\Phi}_{\boldsymbol{p}}^{(\ell_1, \ell_2 , \ell_3)}(z)
    =
    r_{\boldsymbol{p}}^{(\ell_1, \ell_2 , \ell_3)}(z;0)
  \end{equation}
  where we have an analogue of the
  period function
  \begin{equation}
    \label{define_r_function}
    r_{\boldsymbol{p}}^{(\ell_1, \ell_2, \ell_3)}(z ; \alpha)
    =
    \frac{1}{
      \sqrt{2 \, P \, \mathrm{i}}
    } \,
    \int_\alpha^\infty
    \frac{\Phi_{\boldsymbol{p}}^{(\ell_1, \ell_2, \ell_3)}(\tau)}{
      \sqrt{\tau-z}
    } \,
    \mathrm{d} \tau
  \end{equation}
  for $\alpha \in \mathbb{Q}$ and $z\in \mathbb{H}^-$.
  On the other hand,  substituting eq.~\eqref{modular_Phi} for eq.~\eqref{define_Phi_hat},
  we get for $z= x + \mathrm{i} \, y$ with $y < 0$
  as follows;
  \begin{align*}
    \widehat{\Phi}_{\boldsymbol{p}}^{(\ell_1 , \ell_2, \ell_3)}(z)
    & =
    \frac{1}{\sqrt{2 \, P \, \mathrm{i}}}
    \sum_{n \in \mathbb{Z}}
    n \,
    \chi_{2 P}^{(\ell_1 , \ell_2 ,\ell_3)}(n) \,
    \int_{z^*}^\infty
    \frac{
      \mathrm{e}^{\frac{n^2}{2 P} \pi \mathrm{i} \tau}
    }{
      \sqrt{\tau - z}
    } \,
    \mathrm{d} \tau
    \\
    & =
    \sum_{n=0}^{\infty}
    \chi_{2 P}^{(\ell_1 , \ell_2 ,\ell_3)}(n) \,
    \mathrm{e}^{
      \frac{n^2}{2 P} \pi \mathrm{i} z
    } \,
    \erfc
    \left(
      n \,
      \sqrt{
        -\frac{\pi \, y}{P}
      }
    \right)
  \end{align*}
  As a limit of $z \to \alpha \in \mathbb{Q}$, we find that
  \begin{equation}
    \widehat{\Phi}_{\boldsymbol{p}}^{(\ell_1 , \ell_2, \ell_3)}(\alpha)
    =
    \widetilde{\Phi}_{\boldsymbol{p}}^{(\ell_1 , \ell_2, \ell_3)}(\alpha)
  \end{equation}
  Then a nearly modular property~\eqref{asymptotics_Eichler} of
  $ \widetilde{\Phi}_{\boldsymbol{p}}^{(\ell_1 , \ell_2, \ell_3)}(1/N)$
  follows from eq.~\eqref{modular_Phi_hat} by setting $z \to 1/N$ and
  taking an asymptotic expansion of the
  integral~\eqref{define_r_function} in $N\to\infty$.
\end{proof}

It should be noted that
the sum
in  the first
term in the right hand side of eq.~\eqref{asymptotics_Eichler}
runs over $D(p_1,p_2,p_3)$ distinct triples, but, as was clarified in
the preceding sections,
some of them vanish and
there are   $\gamma(p_1,p_2,p_3)$ non-zero contributions,
which correspond to a sum of the flat connection~\eqref{asymptotic_Witten}.

\section{The Witten--Reshetikhin--Turaev Invariant and the Eichler Integral}
\label{sec:WRT}

We have shown that the Eichler integral at $\tau\to N \in \mathbb{Z}$ gives the Chern--Simons
invariant of the Brieskorn homology spheres
$\Sigma(p_1,p_2,p_3)$.
We shall show that
the WRT invariant for the Brieskorn homology spheres can be expressed in terms of 
a limiting value of the Eichler integral at $\tau\to 1/N$ which 
is given in eq.~\eqref{Eichler_fraction}.

\begin{theorem}
  \label{theorem:invariant_Eichler}
  For the Brieskorn homology spheres $\Sigma(p_1,p_2,p_3)$ such that
  \begin{equation*}
    \frac{1}{p_1} +     \frac{1}{p_2}+     \frac{1}{p_3} <1
  \end{equation*}
  we have
  \begin{equation}
    \label{invariant_and_Eichler}
    \mathrm{e}^{
      \frac{2 \pi \mathrm{i}}{N}
      ( \frac{\phi(p_1,p_2,p_3)}{4} - \frac{1}{2} )
    } \,
    \left(
      \mathrm{e}^{\frac{2 \pi \mathrm{i}}{N}} - 1
    \right)
    \,    \tau_N \bigl( \Sigma(p_1, p_2, p_3) \bigr)
    =
    \frac{1}{2} \,
    \widetilde{\Phi}_{\boldsymbol{p}}^{(1,1,1)}(1/N)
  \end{equation}
\end{theorem}

\begin{proof}
  Proof is essentially same with one given in Ref.~\citen{LawrZagi99a}.

  We use the Gauss sum;
  \begin{equation}
    \label{Gauss_sum}
    G(N)=
    \sum_{n=0}^{2  N -1}
    \mathrm{e}^{- \frac{1}{2  N} n^2 \pi \mathrm{i}}
    =
    \sqrt{2 \,  N} \,
    \mathrm{e}^{-\pi \mathrm{i}/4}
  \end{equation}
  We see that
  \begin{align}
    G(N) &=
    \sum_{k=0}^{2  N-1} \mathrm{e}^{
      -\frac{1}{2  N} (k-n)^2 \pi \mathrm{i}
    }
    \nonumber
    \\
    & =
    \mathrm{e}^{-\frac{1}{2  N} n^2 \pi \mathrm{i}} \,
    \sum_{k=0}^{2  N-1}
    \mathrm{e}^{-\frac{1}{2  N} k^2 \pi \mathrm{i}
      + \frac{1}{ N} k n \pi \mathrm{i}
    }
    \label{Gauss_sum_identity}
  \end{align}
  These identities follow from the reciprocity
  formula~\eqref{Gauss_sum_formula}.

  As we see that $\widetilde{\Phi}_{\boldsymbol{p}}^{(1,1,1)}(1/N)$
  is given by eq.~\eqref{Eichler_and_L},
  we have using Prop.~\ref{prop:L_function}
  as follows;
  \begin{align*}
    \widetilde{\Phi}_{\boldsymbol{p}}^{(1,1,1)}(1/N)
    & =
    \lim_{t \searrow 0} \sum_{n=0}^\infty
    \chi_{2 P}^{(1,1,1)}(n) \,
    \mathrm{e}^{\frac{1}{2 P N} n^2 \pi \mathrm{i}}  \,
    \mathrm{e}^{- n t}
    \\
    & =
    \lim_{t \searrow 0}
    \frac{1}{G(P \, N)}
    \sum_{n=0}^\infty \sum_{k=0}^{2 P N-1}
    \chi_{2 P}^{(1,1,1)}(n) \,
    \mathrm{e}^{
      -\frac{1}{2 P N} k^2 \pi \mathrm{i} 
      +
      n \left(
        \frac{k}{P N} \pi \mathrm{i} - t
      \right)
    }
  \end{align*}
  where in the second equality we have applied
  eq.~\eqref{Gauss_sum_identity}.

  We  recall that we have the generating function of the odd periodic
  function $\chi_{2P}^{(1,1,1)}(n)$ as
  \begin{equation}
    \label{generate_chi_general}
    \frac{
      (z^{p_1 p_2} - z^{-p_1 p_2}) \,
      (z^{p_2 p_3} - z^{-p_2 p_3}) \,
      (z^{p_1 p_3} - z^{-p_1 p_3})
    }{
      z^{p_1 p_2 p_3} - z^{-p_1 p_2 p_3}
    }
    =
    \sum_{n=0}^\infty
    \chi_{2 P}^{(1,1,1)}(n) \, z^{n}
  \end{equation}
  Thus we get
  \begin{equation*}
    \widetilde{\Phi}_{\boldsymbol{p}}^{(1,1,1)}(1/N)
    =
    \frac{1}{G(P \, N)}
    \sum_{\substack{
        k=0 \\
      N \nmid k
    }}^{2 P N-1}
    \mathrm{e}^{- \frac{1}{2 P N} k^2 \pi \mathrm{i}} \,
    \frac{
      \prod_{j=1}^3
      \left(
        \mathrm{e}^{\frac{k}{N p_j} \pi \mathrm{i}}
        -
        \mathrm{e}^{-\frac{k}{N p_j} \pi \mathrm{i}}
      \right)
    }{
      \mathrm{e}^{\frac{k}{N} \pi \mathrm{i}}
      -
      \mathrm{e}^{ -\frac{k}{N} \pi \mathrm{i}}
    }
  \end{equation*}
  Here a contribution from the  sum over $N \mid k$ is written by use of
  eq.~\eqref{Gauss_sum_formula} as
  \begin{align*}
    & \lim_{t \searrow 0} \frac{1}{G(P \, N)}
    \sum_{n=0}^\infty \sum_{m \mod 2 P}
    \chi_{2 P}^{(1,1,1)}(n) \,
    \mathrm{e}^{-\frac{N}{2 P} m^2 \pi \mathrm{i}
      +
      \frac{m n}{P} \pi \mathrm{i} - n t
    }
    \\
    & =
    \lim_{t\searrow 0}
    \frac{1}{N}
    \sum_{n=0}^\infty \sum_{k \mod N}
    \chi_{2 P}^{(1,1,1)}(n) \,
    \mathrm{e}^{\frac{2 P}{N} \pi \mathrm{i}
      \left(
        k+\frac{n}{2P}
      \right)^2
      -n t
    }
  \end{align*}
  which vanishes due to the odd periodicity
  $\chi_{2 P}^{(1,1,1)}(n)
  =
  -
  \chi_{2 P}^{(1,1,1)}(2 \, P-n)$.
  Comparing with an expression~\eqref{Rozansky},
  this completes a proof of eq.~\eqref{invariant_and_Eichler}.
\end{proof}

\begin{theorem}[\cite{LawrZagi99a}]
  For the Poincar{\'e} homology sphere 
  with $\boldsymbol{p} =(2,3,5)$,
  we have 
    \begin{equation}
    \mathrm{e}^{
      \frac{2 \pi \mathrm{i}}{N} 
    } \,
    \left(
      \mathrm{e}^{\frac{2 \pi \mathrm{i}}{N}} - 1
    \right)
    \,    \tau_N \bigl( \Sigma(2,3,5) \bigr)
    =
    1
    +
    \frac{1}{2} \,
    \mathrm{e}^{- \frac{1}{60 N} \pi \mathrm{i}}
    \cdot
    \widetilde{\Phi}_{\boldsymbol{p}}^{(1,1,1)}(1/N)
  \end{equation}
\end{theorem}

\begin{proof}
  A proof is same with Theorem~\ref{theorem:invariant_Eichler}
  and was done in Ref.~\citen{LawrZagi99a}.
  A difference follows from a fact that
  the odd periodic function satisfies
  \begin{equation}
    \frac{
      (z^{6} - z^{- 6}) \,
      (z^{10} - z^{- 10}) \,
      (z^{15} - z^{- 15})
    }{
      z^{30} - z^{- 30}
    }
    =
    \frac{1}{z} + z +
    \sum_{n=0}^\infty
    \chi_{60}^{(1,1,1)}(n) \, z^{n}
  \end{equation}
  in place of eq.~\eqref{generate_chi_general}.
\end{proof}

Combining these theorems with a nearly modular property of the Eichler
integral~\eqref{asymptotics_Eichler}, we obtain an asymptotic expansion of the quantum invariant
in $N\to\infty$.

\begin{coro}
  For a case of $\frac{1}{p_1}+\frac{1}{p_2}+\frac{1}{p_3}<1$ 
  we have
  \begin{multline}
    \label{tau_limit}
    \mathrm{e}^{
      \frac{2 \pi \mathrm{i}}{N}
      ( \frac{\phi(p_1,p_2,p_3)}{4} - \frac{1}{2} )
    } \,
    \left(
      \mathrm{e}^{\frac{2 \pi \mathrm{i}}{N}} - 1
    \right)
    \,    \tau_N \bigl( \Sigma(p_1, p_2, p_3) \bigr)
    \\
    \simeq
    \sqrt{\frac{N}{ \mathrm{i}} }
    \sum_{\ell_1, \ell_2, \ell_3}
    \mathbf{S}_{1,1,1}^{\ell_1, \ell_2, \ell_3} \,
    \mathrm{e}^{- \frac{1}{2} \pi \mathrm{i} P N
      \left(
        1+\sum_j \frac{\ell_j}{p_j}
      \right)^2
    }
    + \frac{1}{2} \sum_{k=0}^\infty
    \frac{L(-2 \, k , \chi_{2 P}^{(1,1,1)})}{k!}
    \,
    \left(
      \frac{\pi \, \mathrm{i}}{
        2 \, P \, N
      }
    \right)^k
  \end{multline}
\end{coro}
Here the sum in the first term runs over $\gamma(p_1, p_2, p_3)$
distinct triples, \emph{i.e.}, the triple $(\ell_1, \ell_2, \ell_3)$
satisfies a constraint~\eqref{ell_condition} under a
symmetry~\eqref{ell_symmetry}, and we have used
eq.~\eqref{integer_Chern}.

We give some examples below:
\begin{itemize}
\item $\Sigma(2,3,7)$:

  The function  $\Phi^{(\ell_1, \ell_2, \ell_3)}(\tau)$ spans
  a $D(2,3,7)=3$-dimensional space, and the independent functions
  are given for
  $(\ell_1, \ell_2, \ell_3)=(1,1,1), (1,1,2)$, and $(1,1,3)$.
  For these triples 
  we have  from
  eq.~\eqref{Eichler_integer}
  that
  $\widetilde{\Phi}^{(\ell_1, \ell_2, \ell_3)}(N)
  =0,
  -2 \, \mathrm{e}^{\frac{25}{84} \pi \mathrm{i} N},
  -2 \, \mathrm{e}^{-\frac{47}{84} \pi \mathrm{i} N}
  $,
  which shows 
  $\CS(A)=-\frac{25}{168},
  \frac{47}{168}$.
  Indeed  we see that
  $(\ell_1,\ell_2,\ell_3)=
  (1,1,1)$ does not satisfy a condition~\eqref{ell_condition}.
  This fact shows
  $\gamma(2,3,7)=2$, and
  is consistent with eq.~\eqref{gamma_Casson} as we have
  $\lambda_C\bigl( \Sigma(2,3,7) \bigr)=-1$.
  As a result, we have
  \begin{multline*}
    \mathrm{e}^{
      - \frac{2 \pi \mathrm{i}}{N}
      \frac{167}{168} 
    } \,
    \left(
      \mathrm{e}^{\frac{2 \pi \mathrm{i}}{N}} - 1
    \right)
    \,    \tau_N \bigl( \Sigma(2,3,7) \bigr)
    \\
    \simeq
    \sqrt{\frac{N}{ \mathrm{i} }} \,
    \frac{2}{\sqrt{7}} \,
    \left(
      - \sin \left(     \frac{2 \, \pi }{7}  \right)
      \mathrm{e}^{- \frac{25}{84} \pi \mathrm{i}  N}
      -
      \sin \left(     \frac{3 \, \pi }{7}  \right)
      \mathrm{e}^{ \frac{47}{84} \pi \mathrm{i}  N}
    \right)
    \\
    + \frac{1}{2} \sum_{k=0}^\infty
    \frac{L(-2 \, k , \chi_{84}^{(1,1,1)})}{k!}
    \,
    \left(
      \frac{\pi \, \mathrm{i}}{
        84 \, N
      }
    \right)^k
  \end{multline*}
  where
  \begin{equation*}
    \begin{array}{c||ccccccccc}
      n \mod 84 &
      1 & 13 & 29 & 41 & 43 & 55 & 71 & 83 &
      \text{others}
      \\
      \hline
      \chi_{84}^{(1,1,1)}(n)
      & 1 & -1 &-1 & 1 & -1 & 1 & 1 & -1 & 0
    \end{array}
  \end{equation*}

\item $\Sigma(3,4,5)$:

  There are  $D(3,4,5)=6$ independent functions 
  $  \Phi^{(\ell_1, \ell_2, \ell_3)}(\tau)$;
  $(\ell_1, \ell_2, \ell_3)
  =
  (1,1,1),
  (1,1,2),
  (1,1,3),
  (1,1,4),
  (1,2,1)$,
  and $
  (1,2,2)
  $.
  For these triples, we have the Eichler integral
  $
  \widetilde{\Phi}^{(\ell_1, \ell_2, \ell_3)}(N)
  =
  0,
  0,
  -2 \, \mathrm{e}^{-\frac{119}{120} \pi \mathrm{i} N},
  -2 \, \mathrm{e}^{\frac{49}{120} \pi \mathrm{i} N},
  -2 \, \mathrm{e}^{\frac{1}{30} \pi \mathrm{i} N},
  -2 \, \mathrm{e}^{-\frac{11}{30} \pi \mathrm{i} N}
  $ respectively,
  which indicates
  $\CS(A)
  = \frac{119}{240},
  -\frac{49}{240},
  -\frac{1}{60},
  \frac{11}{60}
 $.
  We can check that
  two triples,  $(1,1,1)$ and $(1,1,2)$,  do not satisfy
  a constraint~\eqref{ell_condition}.
  We thus have $\gamma(3,4,5)=4$, which is consistent with
  $\lambda_C\bigl(\Sigma(3,4,5)\bigr)=-2$.
  Then we obtain
  \begin{multline*}
    \mathrm{e}^{
      - \frac{2 \pi \mathrm{i}}{N}
      \frac{71}{240} 
    } \,
    \left(
      \mathrm{e}^{\frac{2 \pi \mathrm{i}}{N}} - 1
    \right)
    \,    \tau_N \bigl( \Sigma(3,4,5) \bigr)
    \\
    \simeq
    \sqrt{\frac{N}{ \mathrm{i} }} \,
    \frac{1}{\sqrt{5}} \,
    \Biggl(
      - \sin \left(     \frac{ \pi }{5}  \right)
      \mathrm{e}^{ \frac{119}{120} \pi \mathrm{i}  N}
      +
      \sin \left(     \frac{2 \, \pi }{5}  \right)
      \mathrm{e}^{- \frac{49}{120} \pi \mathrm{i}  N}
      \\
      -
      \sqrt{2} \,
      \sin \left(     \frac{2\, \pi }{5} \right)
      \mathrm{e}^{ -\frac{1}{30} \pi \mathrm{i}  N}
      +
      \sqrt{2} \,
      \sin \left(     \frac{ \pi }{5}  \right)
      \mathrm{e}^{ \frac{11}{30} \pi \mathrm{i}  N}
    \Biggr)
    \\
    + \frac{1}{2} \sum_{k=0}^\infty
    \frac{L(-2 \, k , \chi_{120}^{(1,1,1)})}{k!}
    \,
    \left(
      \frac{\pi \, \mathrm{i}}{
        120 \, N
      }
    \right)^k
  \end{multline*}
  where
  \begin{equation*}
    \begin{array}{c||ccccccccc}
      n \mod 120 &
      13 & 37 & 43 & 53 & 67 & 77 & 83 & 107 &
      \text{others}
      \\
      \hline
      \chi_{120}^{(1,1,1)}(n)
      & 1 & -1 &-1 & -1 & 1 & 1 & 1 & -1 & 0
    \end{array}
  \end{equation*}
\end{itemize}

The Poincar{\'e} homology sphere
$\Sigma(2,3,5)$
was studied in Ref.~\citen{LawrZagi99a}.
In this case we have a $D(2,3,5)=2$ dimensional representation of the
modular group $PSL(2;\mathbb{Z})$, and
independent functions
$\widetilde{\Phi}^{(\ell_1,\ell_2,\ell_3)}(\tau)$ can be defined
for
$(\ell_1,\ell_2, \ell_3)=
(1,1,1)$
and $(1,1,2)$.
We can check that  both triples  fulfill a
condition~\eqref{ell_condition}, and that
we have
$\widetilde{\Phi}^{(\ell_1, \ell_2, \ell_3)}(N)
=
  -2 \, \mathrm{e}^{\frac{1}{60} \pi \mathrm{i} N},
  -2 \, \mathrm{e}^{\frac{49}{60} \pi \mathrm{i} N}
$,
which indicates that
the Chern--Simons
invariant is given by
$-\frac{1}{120}$ and $-\frac{49}{120}$ respectively.
As we have $\phi(2,3,5)=\frac{181}{30}$, we obtain an exact asymptotic expansion as follows;
\begin{coro}[\cite{LawrZagi99a}]
  \begin{multline}
    \label{tau_Poincare_limit}
    \mathrm{e}^{
      \frac{2 \pi \mathrm{i}}{N}
      \frac{121}{120}
    } \,
    \left(
      \mathrm{e}^{\frac{2 \pi \mathrm{i}}{N}} - 1
    \right)
    \,    \tau_N \bigl( \Sigma(2,3,5) \bigr)
    \\
    \simeq
    \sqrt{\frac{N}{ \mathrm{i} }} \,
    \frac{2}{\sqrt{5}} \,
    \left(
      \sin \Bigl( \frac{\pi}{5}  \Bigr) \,
      \mathrm{e}^{- \frac{1}{60} \pi \mathrm{i} N}
      +
      \sin \Bigl( \frac{2 \, \pi}{5}  \Bigr) \,
      \mathrm{e}^{- \frac{49}{60} \pi \mathrm{i} N}
    \right)
    \\
    + \mathrm{e}^{\frac{\pi \mathrm{i}}{60 N} }
    +
    \frac{1}{2}
    \sum_{k=0}^\infty
    \frac{L(-2 \, k , \chi_{60}^{(1,1,1)})}{k!} \,
    \left(
      \frac{\pi \, \mathrm{i}}{60 \, N}
    \right)^k  
  \end{multline}
  where
  \begin{equation*}
    \begin{array}{c||ccccccccc}
      n \mod 60 &
      1 & 11 & 19 & 29 & 31 & 41 & 49 & 59 &
      \text{others}
      \\
      \hline
      \chi_{60}^{(1,1,1)}(n)
      & -1 & -1 &-1 & -1 & 1 & 1 & 1 & 1 & 0
    \end{array}
  \end{equation*}
\end{coro}

\section{
  $\mathbf{S}$-matrix and the Reidemeister Torsion}
\label{sec:torsion}

An asymptotic behavior of the Witten invariant $Z_k(\mathcal{M})$ can
be given from the definition~\eqref{Witten_and_tau}.
In  a large $N$ limit, the first term in eqs.~\eqref{tau_limit}
and~\eqref{tau_Poincare_limit}  dominate an asymptotic behavior of the
quantum invariant, which will be shown to denote a contribution from 
flat connections.
We obtain the following;
\begin{coro}
  We have an asymptotic behavior of the Witten invariant~\eqref{Witten}
  for the Brieskorn homology spheres in $N\to\infty$
  as
  \begin{multline}
    Z_{N-2}\bigl( \Sigma(p_1,p_2, p_3) \bigr)
    \\
    \sim
    \frac{1}{2} \, \mathrm{e}^{-\frac{3}{4} \pi \mathrm{i}} \,
    \mathrm{e}^{-\frac{\phi(p_1,p_2,p_3)}{2 N} \pi \mathrm{i}}
    \,
    \sum_{\ell_1, \ell_2, \ell_3}
    \Bigl(
    \sqrt{2} \, \mathbf{S}_{1,1,1}^{\ell_1, \ell_2, \ell_3}
    \Bigr) \,
    \mathrm{e}^{-\frac{P}{2}
      \left(
        1+ \sum_{j=1}^3 \frac{\ell_j}{p_j}
      \right)^2 
      \pi \mathrm{i} N
    }
  \end{multline}
  where the $\mathbf{S}$-matrix is defined in
  eq.~\eqref{S_transformation}, and the sum  runs over
  $\gamma(p_1,p_2,p_3)$ distinct triples satisfying eq.~\eqref{ell_condition}.
\end{coro}

We see that,
except a decaying  factor
$\mathrm{e}^{-\frac{\phi(p_1,p_2,p_3)}{2 N} \pi  \mathrm{i}}$,
our result proves eq.~\eqref{asymptotic_Witten} exactly
as we have seen that  a dominating
exponential factor denotes the Chern--Simons
invariant~\eqref{CS_B} for the Brieskorn homology spheres.

In fact
we can establish  a relationship 
among the $\mathbf{S}$-matrix, the
Reidemeister torsion, and the spectral flow;
\begin{theorem}
  \begin{equation}
    \label{S_and_torsion}
    \sqrt{2} \,
    \mathbf{S}_{1,1,1}^{\ell_1, \ell_2, \ell_3}
    =
    \sqrt{T_\alpha} \,
    \mathrm{e}^{-2 \pi \mathrm{i}  I_\alpha/4}
  \end{equation}
  where the triple $(\ell_1,\ell_2,\ell_3)$ satisfies eq.~\eqref{ell_condition}.
\end{theorem}

\begin{proof}
A proof of the absolute value of
$\sqrt{2} \,
\mathbf{S}_{1,1,1}^{\ell_1,\ell_2,\ell_3}$
is straightforward since
it is known~\cite{DFreed92a} that
the Reidemeister torsion of the
Brieskorn homology sphere is given by
\begin{equation}
  \sqrt{T_\alpha}
  =
  \frac{8}{\sqrt{P}} \,
  \prod_{j=1}^3
  \left|
    \sin \Bigl(
    P \,
    \frac{\ell_j}{p_j^{~2}}
    \Bigr)
  \right|
\end{equation}

To prove a part of the phase factor,
we recall that
the spectral flow of the Brieskorn homology spheres is given
by~\cite{FintuStern90a}
\begin{multline}
  I_\alpha
  =
  -3 - 
  \Biggl(
    \frac{2 \,
      \bigl(e (\ell_1, \ell_2, \ell_3) \bigr)^2
    }{P}
    \\
    +
    \sum_{j=1}^3
    \frac{2}{p_j} \,
    \sum_{k=1}^{p_j-1}
    \cot
    \Bigl( \frac{k \, P  \, \pi}{p_j^{~2}} \Bigr) \,
    \cot
    \Bigl( \frac{k \, \pi}{p_j} \Bigr) \,
    \sin^2
    \Bigl( \frac{k \,
      e(\ell_1, \ell_2, \ell_3)
      \, \pi}{p_j} \Bigr) 
  \Biggr)
  \mod 8
  \label{spectral}
\end{multline}
where
\begin{equation}
  e \equiv
  e(\ell_1,\ell_2, \ell_3)
  = P \sum_{j=1}^3 \frac{p_j - \ell_j}{p_j}
\end{equation}
Using this we have
\begin{align*}
  \mathrm{e}^{-2 \pi \mathrm{i} I_\alpha/4}
  & =
  \mathrm{e}^{\frac{\pi \mathrm{i}}{2}
    \left( 3 + 2 \frac{e^2}{P} \right)
  } \,
  \prod_{j=1}^3 \exp
  \left(
    \frac{\pi \mathrm{i}}{2} \cdot
    \frac{2}{p_j}
    \sum_{k=1}^{p_j-1}
    \cot
    \left(
      \frac{k \, P \, \pi}{p_j^{~2}}  
    \right)
    \cot
    \left(
      \frac{k \, \pi}{p_j}
    \right)
    \sin^2
    \left(
      \frac{k \, e \, \pi}{p_j} 
    \right)
  \right)
  \\
  & =
  \mathrm{e}^{\pi \mathrm{i} e}
  \,
  \prod_{j=1}^3
  \sign
  \left(
    \sin \left( \frac{q_j \, e \, \pi}{p_j}  \right)
    \,
    \sin \left( \frac{ e \, \pi}{p_j}  \right)
  \right)
\end{align*}
Here in the second equality we have used
$\frac{P}{p_j} \cdot q_j =1 \pmod {p_j}$,
and
an identity~\cite{LCJeff92a,Rozan95a},
\begin{multline}
  -\mathrm{i} \,
  \sign
  \left(
    \sin \left(\frac{r \, n \, \pi}{p}\right) \,
    \sin \left(\frac{n \, \pi}{p}      \right)
  \right) \,
  \mathrm{e}^{\pi \mathrm{i} n}
  \\
  =
  \exp\left(
    -\frac{\pi \mathrm{i}}{2} \,
    \left(
      -2 + \frac{2 \, r \, n^2}{p}
      +
      \frac{2}{p}
      \sum_{k=1}^{p-1}
      \cot \left(\frac{k \, q \, \pi}{p}\right) \,
      \cot \left(\frac{k \, \pi}{p}\right) \,
      \sin^2 \left(\frac{k \, n \, \pi}{p}\right) 
    \right)
  \right)
\end{multline}
where we suppose  $n\in\mathbb{Z}$ and $q \, r =1 \pmod p$.
We further see that
\begin{equation*}
  \prod_{j=1}^3 
  \sin \left(
    \frac{ e \, \pi}{p_j}
  \right)
  =
  (-1)^{1 + 
    \sum_{j <k}
    \left(
      p_j p_k + \ell_j p_k + p_j \ell_k
    \right)
  }
  \prod_{j=1}^3
  \sin\left(
    \pi \, P \, \frac{\ell_j}{p_j^{~2}}
  \right)
\end{equation*}
and
\begin{align*}
  \prod_{j=1}^3 
  \sign \left(
    \sin \left(
      \frac{q_j \, e \, \pi}{p_j}
    \right)
  \right)
  & =
  1
\end{align*}
which follows from
\begin{align*}
  \sin \left( \frac{ q_1 \, e \, \pi}{p_1} \right)
  & =
  (-1)^{1 + q_1 ( p_2 p_3 + \ell_2 p_3 + p_2\ell_3)} 
  \sin\left(
    \pi \, \frac{\ell_1}{p_1}
    \cdot  q_1  \, \frac{P}{p_1} 
  \right)
  \\
  & =
  (-1)^{1+  q_1 ( p_2 p_3 + \ell_2 p_3 + p_2\ell_3) 
    -\ell_1 (p_2 q_3 + q_2 p_3)} \,
  \sin
  \left(
    \frac{\ell_1 \, \pi}{p_1}
  \right)
\end{align*}
Collecting these results, we obtain
\begin{equation*}
  \mathrm{e}^{-2 \pi \mathrm{i} I_\alpha/4}
  =
  (-1)^{
    1+ P + P \sum_{j=1}^3 \frac{\ell_j}{p_j}
    +
    \sum_{j<k}
    \left(
      p_j p_k + \ell_j p_k + p_j \ell_k
    \right)
  } \,
  \prod_{j=1}^3
  \sign
  \left(
    \sin
    \left(
      \frac{ P \, \ell_j \, \pi}{p_j^{~2}}
    \right)
  \right)
\end{equation*}
which proves a phase factor of eq.~\eqref{S_and_torsion}.
\end{proof}

We  reconsider some  examples from  section~\ref{sec:WRT};
\begin{itemize}
\item the Poincar{\'e} homology sphere $\Sigma(2,3,5)$:

  We have two triples
  $(\ell_1, \ell_2, \ell_3)
  =(1,1,1)$ and $ (1,1,2)$.
  These respectively
  give the spectral flow~\eqref{spectral}
  $  I_\alpha = 4$ and
  $ 0$,
  which supports eq.~\eqref{tau_Poincare_limit}.

\item $\Sigma(2,3,7)$:

  We have
  $(\ell_1, \ell_2, \ell_3)
  =
  (1,1,2)$
  and $(1,1,3)$,
  which give
  $  I_\alpha = 6$ and $2$ respectively.
  This is consistent with a result in previous section.

\item $\Sigma(3,4,5)$:

  We have four triples
  $(\ell_1, \ell_2, \ell_3)
  =(1,1,3),
  (1,1,4),
  (1,2,1)$, and
  $(1,2,2)$.
  For these irreducible representations, we get
  from eq.~\eqref{spectral} 
  $  I_\alpha = 2, 4, 6,$ and
  $ 0$.
  This result is consistent with an asymptotic expansion given before.
\end{itemize}
\section{The Ohtsuki Invariant}
\label{sec:Ohtsuki}

{}From  asymptotic expansions~\eqref{tau_limit}
and~\eqref{tau_Poincare_limit} of the WRT invariant,
we can introduce a formal power
series which is  ignored in section~\ref{sec:torsion}.
This series may denote a trivial connection contribution~\cite{LRozan96f,LRozan97a,LawreRozan99a}.
By regarding $\mathrm{e}^{2 \pi \mathrm{i}/N}$ as $q$,
we can define $\tau_\infty(\mathcal{M})$
for a case of
$\frac{1}{p_1} + \frac{1}{p_2} +\frac{1}{p_3} <1$ 
\begin{equation}
  \label{infinity_general}
  q^{
    \frac{\phi(p_1,p_2,p_3)}{4} - \frac{1}{2} 
  } \,
  \left(
    q
    - 1
  \right)
  \,
  \tau_\infty\bigl( \Sigma(p_1, p_2, p_3) \bigr)
  =
  \frac{1}{2} \sum_{k=0}^\infty
  \frac{L(-2 \, k , \chi_{2 P}^{(1,1,1)})}{k!}
  \,
  \left(
    \frac{\log q}{
      4 \, P 
    }
  \right)^k
\end{equation}
and for the Poincar{\'e} homology sphere
\begin{equation}
  \label{infinity_Poincare}
  q^{
    \frac{121}{120}
  } \,
  \left(
    q    - 1
  \right)
  \,    \tau_\infty \bigl( \Sigma(2,3,5) \bigr)
  =
  q^{\frac{1}{120} }
  +
  \frac{1}{2}
  \sum_{k=0}^\infty
  \frac{L(-2 \, k , \chi_{60}^{(1,1,1)})}{k!} \,
  \left(
    \frac{\log q}{120}
  \right)^k  
\end{equation}
With these definitions
the Ohtsuki invariant~\cite{TOhtsu96a} is
defined 
by
the formal series for
$\tau_\infty(\mathcal{M})$;
\begin{equation}
  \tau_\infty\bigl( \Sigma(p_1, p_2, p_3) \bigr)
  =
  \sum_{n=0}^\infty
  \lambda_n \bigl( \Sigma(p_1, p_2, p_3) \bigr)
  \cdot (q-1)^{n}
\end{equation}

Infinite series in eqs.~\eqref{infinity_general}
and~\eqref{infinity_Poincare}
originate from an asymptotic expansion of the integral
$r_{\boldsymbol{p}}^{(1,1,1)}(1/N ; 0)$ defined in
eq.~\eqref{define_r_function}, which appears as a \emph{tail} of the
nearly modular property of the Eichler integral.
It is noted that an integral formula for the Ohtsuki invariant was studied in Refs.~\citen{LRozan97a,RLawren99a}
by a different method.

To compute an explicit form of $\lambda_n(\mathcal{M})$,
we use the  Stirling number of the first kind defined by
\begin{equation}
  \prod_{j=0}^{n-1}(x - j)
  =
  \sum_{m=0}^n
  S_n^{(m)}  \, x^m
\end{equation}
As  the Stirling number satisfies
(see, \emph{e.g.}, Ref.~\citen{AbraSteg72Book})
\begin{equation}
  \frac{(\log q)^m}{m!}
  =
  \sum_{n=m}^\infty
  S_n^{(m)} \,
  \frac{ (q -1 )^n }{n!}
\end{equation}
we can easily obtain the following expression.

\begin{theorem}
  Let the function $\Lambda_n(p_1,p_2,p_3)$ be defined by
  \begin{multline}
    \Lambda_{n}(p_1,p_2,p_3)
    =
    \frac{1}{2} \, \frac{1}{(n+1)!} \,
    \sum_{m=1}^{n+1}
    S_{n+1}^{(m)} \,
    \left(
      \frac{2-\phi(p_1,p_2,p_3)}{4} 
    \right)^m
    \\
    \times
    \sum_{k=0}^m
    \begin{pmatrix}
      m \\
      k
    \end{pmatrix}
    \, 
    \left(
      \frac{1}{P \, (2 - \phi(p_1,p_2,p_3))}
    \right)^k \,
    L(-2 \, k , \chi_{2 P}^{(1,1,1)})
  \end{multline}
  Then the invariant $\lambda_n(\mathcal{M})$ is computed as follows;
  \begin{equation}
    \label{lambda_Brieskorn}
    \lambda_{n}\bigl( \Sigma(p_1, p_2, p_3) \bigr)
    =
    \begin{cases}
      \Lambda_n( p_1,p_2, p_3)
      &
      \text{for $\displaystyle
        \frac{1}{p_1} +     \frac{1}{p_2}+    \frac{1}{p_3} <1
        $}
      \\[5mm]
      \Lambda_n(2,3,5) + (-1)^{n+1}
      &
      \text{for the Poincar{\'e} homology sphere}
    \end{cases}
  \end{equation}
\end{theorem}

We note that
a value of the $L$-function, which is given by eq.~\eqref{L_and_Bernoulli},
can be computed easily from a generating function for  a case of
$(\ell_1,\ell_2,\ell_3)=(1,1,1)$;
\begin{align}
  -2 \, \ch (z)
  +
  2 \,
  \frac{
    \sh ( 6 \, z) \,
    \sh ( 10 \, z) \,
    }{
      \ch ( 15 \, z)
    }
    & =
    - 2 \,
    \frac{\ch (9 \, z) \, \ch(5 \, z)}{
      \ch(15 \, z)
    }
    \nonumber
    \\
    & =
    \sum_{n=0}^\infty
    \frac{
      L(-2 \, n , \chi_{60}^{(1,1,1)})
    }{(2 \, n)!} \,
    z^{2 n}
  \end{align}
and in a case of
$
\frac{1}{p_1} +     \frac{1}{p_2}+     \frac{1}{p_3} <1
$ we have
\begin{equation}
  \label{generating_Sigma}
  4 \,
  \frac{
    \sh ( p_1 \, p_2 \, z) \,
    \sh ( p_1 \, p_3 \, z) \,
    \sh ( p_2 \, p_3 \, z) 
    }{
      \sh ( p_1 \, p_2 \, p_3 \, z)
    }
    =
    \sum_{n=0}^\infty
    \frac{
      L(-2 \, n , \chi_{2 P}^{(1,1,1)})
    }{(2 \, n)!} \,
    z^{2 n}
\end{equation}

See Table~\ref{table:lambda} for explicit values of
$\lambda_n(\mathcal{M})$ for some Brieskorn spheres.

\begin{table}[]
  \rotatebox[]{90}{
    \footnotesize
    \begin{minipage}{23cm
}
      \begin{center}
        \begin{equation*}
          \begin{array}{c|rrrrrrrrr}
            \mathcal{M} & \lambda_0 & \lambda_1 &
            \lambda_2 & \lambda_3 & \lambda_4 & \lambda_5 & \lambda_6
            & \lambda_7 & \lambda_8
            \\
            \hline \hline
            \Sigma(2,3,5)
            & 1 & -6 & 45 &  -464 & 6224 & - 102816 & 2015237 & -45679349
            & 1175123730
            \\
            \Sigma(2,3,7)
            &  1 & -6 & 69 & -1064 & 20770 & -492052 & 13724452 & -440706098
            & 16015171303
            \\
            \Sigma(2,3,11)
            &  1 & -12 & 198 & -4564 & 136135 & -4979568 & 215636785
            & -10785653847 & 611802510704
            \\
            \Sigma(2,3,13)
            & 1 & -12 & 246 & -6916 & 248171 & -10848488 & 559466999
            & -33256127501 & 2238888918356
            \\
            \Sigma(2,3,17)
            &  1 &-18 & 459 & -16404 & 757689 & -42883758 & 2872307319
            & -222158381412 & 19483805436567
            \\
            \Sigma(2,3,19)
            & 1 & -18 & 531 & -21660 & 1131375 & -72097914 &
            5424649644
            & -470672677647 & 46266101270760
            \\
            \Sigma(2,5,7)
            &  1 & -12 & 222 & -5596 & 179985 & -7054432 & 326278974 &
            -17397305298 & 1050720467092
            \\
            \Sigma(2,5,9)
            &  1 & -18 & 411 & -12900 & 523445 & -26063974 & 1537243785 &
            -104755839122 & 8097415424747
            \\
            \Sigma(2,5,11)
            & 1 & -18 & 531 & -21180 & 1074975 & -66390674 & 4834590264
            & -405657513711 & 38541528405358
            \\
            \Sigma(2,7,13)
            &  1 & -36 & 1674 & -106884 & 8799855 & -887883368 & 106042462590
            & -14627548503126 & 2288188525438231
            \\
            \Sigma(2,7,15)
            & 1 & -36 & 2010 & -152244 & 14703739 & -1730017752 &
            240158450652 &  -38429255864768 & 6964744996791857
            \\
            \Sigma(2,11,17)
            &  1 & -72 & 6948 & -918744 & 156141090 & -32466056280 &
            7983578235864
            & -2266232536578132 & 729278178446689719
            \\
            \Sigma(2,11,19)
            & 1 & -78 & 8481 & -1261160 & 240437790 & -56007699396 &
            15418103295783
            & -4897366295772501 & 1763003584636961535
            \\
            \Sigma(2,11,21)
            & 1 & -90 & 10695 & -1742580 & 365286685 & -93734684478 &
            28454533630530
            & -9972833783229875 & 3962937841176563555
            \\
            \Sigma(3,4,5)
            &  1 & -12 & 198 & -4324 & 119455 & -4012828 & 159008935 &
            -7263759799 & 375878922067
            \\
            \Sigma(3,4,7)
            &  1 & -18 & 411 & -12420 & 476645 & -22300734 & 1232660885 &
            -78624211186 & 5684458291305
            \\
            \Sigma(3,4,11)
            & 1 & -30 & 1065 & -49960 & 2988770 & -218577416 & 18915594545
            & -1890831231245 & 214380047624390
            \\
            \Sigma(3,4,13)
            & 1 & -30 & 1305 & -74680 & 5390950 & -472424616 & 48786184083
            & -5804487459615 &781992244219680
            \\
            \Sigma(3,4,23)
            & 1 & -60 & 4470 & -441740 & 55544845 & -8523297832 &
            1546017437658 & -323709524734970 & 76844211062714480
            \\
            \Sigma(3,4,25)
            & 1 & -60 & 4950 & -540140 & 74583605 & -12524972472 &
            2481232929734 & -566688382409942 & 146614680355157664
            \\
            \Sigma(3,5,7)
            &  1 & -24 & 684 & -25640 & 1222766 & -71219336 & 4906476652 &
            -390356879176 & 35220329064877
            \\
            \Sigma(3,5,8)
            & 1 & -24 & 804 & -35360 & 1961016 & -132053796 &
            10481576931 & -958711945083 & 99307064129868
            \\
            \Sigma(3,5,11)
            & 1 & -36 & 1626 & -96756 & 7300091 & -671133288 &
            72881361140 & -9132371657296 & 1297049789194653
            \\
            \Sigma(3,5,13)
            & 1 & -42 & 2247 & -158396 & 14146195 & -1538635378 &
            197615046741 & -29280417770120 & 4916816674605230
            \\
            \Sigma(4,5,7)
            &  1 & -30 & 1185 & -60880 & 3934980 & -308750176 & 28560576934 &
            -3045183200982 & 367764306523118
            \\
            \Sigma(4,7,9)
            &  1 & -60 & 4230 & -385820 & 44419865 & -6227423152 &
            1031445122700 &  -197181302142540 & 42735797918044660
          \end{array}
        \end{equation*}
      \end{center}
      \caption{$\lambda_n \bigl( \Sigma(p_1, p_2, p_3) \bigr) $}
      \label{table:lambda}
    \end{minipage}
  }
\end{table}

We can rewrite eq.~\eqref{lambda_Brieskorn} for the first three terms
as follows;
\begin{align}
  \lambda_0 \bigl( \Sigma(p_1, p_2, p_3) \bigr)
  & = 1
  \\[2mm]
  \lambda_1 \bigl( \Sigma(p_1, p_2, p_3) \bigr)
  & =
  -\frac{1}{4} \,
  \left(
    \phi + P \,
    \Bigl(
      1-\frac{1}{p_1^{~2}}-\frac{1}{p_2^{~2}}-\frac{1}{p_3^{~2}}
    \Bigr)
  \right)
  \\[2mm]
  \lambda_2 \bigl( \Sigma(p_1, p_2, p_3) \bigr)
  & =
  \frac{1}{12} \,\Biggl(
    \frac{3 \, \phi^2 + 12 \, \phi -4}{8}
    +
    \frac{3 \, P}{4} \, (\phi +2) \,
    \Bigl(
      1- \sum_{j=1}^3 \frac{1}{p_j^{~2}}
    \Bigr)
    \nonumber 
    \\
    & \qquad \qquad 
    +
    \frac{P^2}{8} \,
    \left(
      2 \, \Bigl( 1- \sum_{j=1}^3 \frac{1}{p_j^{~4}} \Bigr)
      +
      5 \, \Bigl( 1- \sum_{j=1}^3 \frac{1}{p_j^{~2}} \Bigr)^2
    \right)
  \Biggr)
\end{align}
where we mean $\phi=\phi(p_1, p_2, p_3)$ defined in eq.~\eqref{phi_definition}.
As was proved  in Refs.~\citen{HMuraka93a,HMuraka95a} we see that
\begin{gather}
  \lambda_0
  \bigl( \Sigma(p_1, p_2, p_3) \bigr)
  =
  1
  \\[2mm]
  \lambda_1
  \bigl( \Sigma(p_1, p_2, p_3) \bigr)
  =
  6 \,
  \lambda_C  \bigl( \Sigma(p_1, p_2, p_3) \bigr)
\end{gather}
where
$\lambda_C(\mathcal{M})$ is the Casson invariant~\eqref{Casson_sphere}
for $\mathcal{M}$.
It is noted that
$\lambda_2\bigl(\Sigma(p_1,p_2,p_3)\bigr)$
is calculated  in Ref.~\citen{CSato97a}.

\section{Discussions}

We have studied the Witten--Reshetikhin--Turaev
invariant for the Brieskorn homology spheres~\eqref{Rozansky}
by use of properties of the Eichler integral of the half-integral weight
based on the  method of Ref.~\citen{LawrZagi99a}.
The WRT invariant coincides with
a limiting value of the Eichler integral at $\tau\to 1/N$ for $N\in\mathbb{Z}$,
and
the nearly modular property~\eqref{asymptotics_Eichler}
of the Eichler integral gives an exact asymptotic
behavior of the WRT invariant~\eqref{tau_limit} and~\eqref{tau_Poincare_limit}.
With the correspondence between the modular form and the quantum invariant,
we can give an interpretation
for the  invariants of manifold
such as the Chern--Simons invariant,
the Ohtsuki invariant,
the Casson invariant,
Reidemeister torsion, 
and the spectral flow
from a point of view of the modular form.
Especially the number of the non-vanishing Eichler integrals
at $\tau\to N\in\mathbb{Z}$
is related to the Casson invariant, and there
exists a correspondence with an irreducible 
SU(2) representation of the fundamental group.
In our previous papers~\cite{KHikami03c,KHikami03a,KHikami03b},
we revealed a relationship between
a specific value of the colored Jones polynomial for the torus knot and link,
$\mathcal{T}_{s,t}$ and $\mathcal{T}_{2,2 m}$,
and the Eichler integral of the
half-integral weight modular form.
Therein shown was that  an exact asymptotic behavior has a form of eqs.~\eqref{tau_limit} and~\eqref{tau_Poincare_limit},
and that
a generating function of a tail polynomial part,
or the Ohtsuki invariant,
is an inverse of the Alexander polynomial
$\frac{A^{1/2}- A^{-1/2}}{\Delta(A)}$.
So we may conclude that
the left hand side of eq.~\eqref{generating_Sigma} plays a role of the inverse of the Alexander polynomial.
In the same manner, 
we may define
an analogue of the Casson invariant 
by a minus half of  the number of the non-vanishing Eichler integrals at $\tau\in \mathbb{Z}$.
We collect these correspondence
in the SU(2) quantum invariants in Table~\ref{table:Alexander}.

\begin{table}[htbp]
  \centering
  \begin{equation*}
    \begin{array}{c|ccc}
      &
      \text{torus link $\mathcal{T}_{2, 2 m }$}
      &
      \text{torus knot $\mathcal{T}_{s,t}$}
      & \Sigma(p_1,p_2,p_3)
      \\[2mm]
      \hline \hline
      \\
%
%
      d
      & 1 & 2 & 3
      \\[2mm]
      \text{``Casson'' $\lambda_C$}
      &
      \displaystyle
      -\frac{1}{2}\, (m-1)
      &
      \displaystyle
      -\frac{1}{4} \, (s-1) \, (t-1)
      &
      \displaystyle
      -\frac{1}{2} \, \gamma(p_1, p_2, p_3)
      \\[6mm]
      \text{``Alexander'' $\Delta(A)$}
      &
      \displaystyle
      \frac{A^m - A^{-m}}{A^{\frac{1}{2}} + A^{-\frac{1}{2}}}
      &
      \displaystyle
      \frac{
        (A^{\frac{s t}{2}} - A^{-\frac{s t}{2}}) \,
        (A^{\frac{1}{2}} - A^{-\frac{1}{2}})
      }{
        (A^{\frac{s}{2}} - A^{-\frac{s}{2}}) \,
        (A^{\frac{t}{2}} - A^{-\frac{t}{2}})
      }
      &
      \displaystyle
      \frac{
        (A^{\frac{1}{2}} - A^{-\frac{1}{2}})^2
      }{
        \prod_{j=1}^3
        (A^{\frac{1}{2 p_j}}- A^{-\frac{1}{2 p_j}})
      }
    \end{array}
  \end{equation*}
  \caption{
    We give an interpretation for the 
    ``Casson invariant'' and the ``Alexander polynomial''
    from the view point of the modular forms.
    ``Dimension'' $d$ denotes that the number of the modular form
    is related to the number of the integral lattice point
    in $d$-dimensional  space.
  }
  \label{table:Alexander}
\end{table}

In Ref.~\citen{LawrZagi99a}
discussed 
also is a relationship between the
WRT invariant for the Poincar{\'e} homology sphere and the Ramanujan mock
theta function.
We hope to report on the $q$-series identity
associated with the quantum invariant for
the Brieskorn homology sphere using a surgery description
with  an expression of the colored
Jones polynomial for the torus knot given in Ref.~\citen{KHikami04a}
(see also Ref.~\citen{TQLe03a}).

\section*{Acknowledgments}
The author would like to thank H.~Murakami for useful discussions
and encouragements.
This work is supported in part  by Grant-in-Aid for Young Scientists
from the Ministry of Education, Culture, Sports, Science and
Technology of Japan.

\bibliographystyle{alphaKH}
\bibliography{_def,gravity,square,math,ba,tba,math5,vm,square2,math4,qalg,math3,math2,poisson,geometry,soliton,cft,knot,tqft,comb,number}

\end{document}